\documentclass{aa}
\usepackage{txfonts}
\usepackage{graphicx}
\usepackage{lscape}
\usepackage{xcolor}

\usepackage{natbib,twoopt}
\usepackage[breaklinks=true]{hyperref} 
\bibpunct{(}{)}{;}{a}{}{,}             
\makeatletter
  \newcommandtwoopt{\citeads}[3][][]{\href{http://adsabs.harvard.edu/abs/#3}%
    {\def\hyper@linkstart##1##2{}%
     \let\hyper@linkend\@empty\citealp[#1][#2]{#3}}}
  \newcommandtwoopt{\citepads}[3][][]{\href{http://adsabs.harvard.edu/abs/#3}%
    {\def\hyper@linkstart##1##2{}%
     \let\hyper@linkend\@empty\citep[#1][#2]{#3}}}
  \newcommandtwoopt{\citetads}[3][][]{\href{http://adsabs.harvard.edu/abs/#3}%
    {\def\hyper@linkstart##1##2{}%
     \let\hyper@linkend\@empty\citet[#1][#2]{#3}}}
  \newcommandtwoopt{\citeyearads}[3][][]%
    {\href{http://adsabs.harvard.edu/abs/#3}
    {\def\hyper@linkstart##1##2{}%
     \let\hyper@linkend\@empty\citeyear[#1][#2]{#3}}}
\makeatother

\usepackage[normalem]{ulem}

\usepackage{xspace}
\usepackage{placeins}

\begin{document} 

\title{Mechanisms for magnetic braking boost and disruption: the role of irradiation-driven winds and convective turnover time spike in cataclysmic variables}

   \author{Vladislav Dodon\inst{1,2}\corrauth{vidodon@kpfu.ru},
           Xiang-Dong Li\inst{1}\email{lixd@nju.edu.cn},
           Xiao-jie Xu\inst{1}\email{xuxj@nju.edu.cn},
           Ilkham Galiullin\inst{2}\email{ilhigaliullin@kpfu.ru}
           \and Askar Sibgatullin\inst{2}\email{absibgatullin@kpfu.ru}
          }
   \institute{School of Astronomy and Space Science, Nanjing University, Nanjing, 210023, People's Republic of China
        \and Kazan Federal University, Kremlevskaya Str.18, 420008, Kazan, Russia \\
        }

\authorrunning{Vladislav Dodon et al.}

   \date{Received XXX; accepted XXX}

\abstract
     {The saturated, boosted, and disrupted magnetic braking (SBD MB) model is an empirical prescription that has recently gained support from observations of diverse close binary systems. Different boosting ($K$) and disruption ($\eta$) parameters appear necessary for different systems, but their physical origins remain uncertain.}
     {We aim to identify the physical mechanisms that boost magnetic braking (MB) and cause its disruption at the fully convective boundary in cataclysmic variables (CVs).}
     {We modelled CV evolution using the MESA code and compared the results with observed CV properties. We computed the convective turnover time ($\tau_c$) directly from the donor's structure rather than adopting empirical relations. We also included irradiation from the accreting white dwarf, which heats the donor's outer layers and can drive additional winds that enhance MB.}
     {The structure-based $\tau_c$ calculation reveals a pronounced spike as the donor approaches full convection, which drives the disruption parameter $\eta$ and initiates the period gap in CVs. The outcome of irradiation is sensitive to the accretion, irradiation, and wind efficiencies, as well as to the base wind mass-loss rate of M-dwarf donors, all of which are poorly constrained from observations. Despite these uncertainties, plausible parameter choices allow irradiation-driven winds to provide the required boost $K$ during accreting phases. We refer to the combined prescription as the i$\tau$SBD MB model and find that it yields evolutionary tracks broadly consistent with the main CV properties.}
     {Our i$\tau$SBD MB framework offers a physically motivated interpretation of the empirical boost and disruption factors in SBD MB for CV evolution. We suggest that the convective turnover time spike at the fully convective boundary may be the universal driver of MB disruption for fast-rotating stars in the saturated regime, while irradiation-driven winds may be the dominant mechanism boosting MB in accreting binaries and other strongly irradiated close systems.}
     

   \titlerunning{Physical SBD magnetic braking in CVs} 
   \maketitle
   \nolinenumbers
%

\section{Introduction}

Magnetic braking (MB) is the principal angular momentum loss (AML) mechanism that governs the spin-down of low-mass single stars and the secular evolution of many close binaries \citep{1962AnAp...25...18S, 1967ApJ...148..217W, 1968MNRAS.138..359M}. In essence, the stellar magnetic field forces its outflowing wind to co-rotate out to the Alfven radius, so the escaping matter removes angular momentum with a much larger lever arm. The predictable spin-down produced by MB forms the basis of gyrochronology, which uses stellar rotation as an age indicator \citep{1972ApJ...171..565S, 2003ApJ...586..464B, 2007ApJ...669.1167B}. In tidally locked binary stars, the torque is removed from the orbital angular momentum, causing the binary to shrink. As a result, MB is efficient in bringing detached binaries into contact, thereby giving rise to accretion-powered systems such as low-mass X-ray binaries (LMXBs) and cataclysmic variables (CVs) \citep{1981A&A...100L...7V, 1983ApJ...275..713R}.

Despite its simple physical picture, MB is a complex process as it depends on the wind mass-loss rate, magnetic field strength and topology, and the dynamo's response to rotation and convection, all of which vary across spectral type and evolutionary state \citep{2012ApJ...754L..26M, 2015ApJ...798..116R, 2015ApJ...799L..23M}. Historically, different MB prescriptions have been developed for particular stellar regimes or observational constraints (e.g. \citealt{1981A&A...100L...7V, 1988ApJ...333..236K, 2003ApJ...582..358A, 2018ApJ...862...90G, 2019ApJ...886L..31V}). In particular, the prescription of \citet{1983ApJ...275..713R} (hereafter RVJ) was adopted extensively in binary evolution studies and underpinned many standard evolutionary models of CVs and LMXBs \citep{1999A&A...350..928T, 2001ApJ...550..897H, 2002ApJ...565.1107P}. When extrapolated beyond their calibration domain, MB prescriptions often disagree by orders of magnitude (e.g. \citealt{2011ApJS..194...28K, 2021ApJ...909..174D, 2022MNRAS.517.4916E, 2023ApJ...950...27G, 2026A&A...707A..76Z}). This lack of universality poses a major obstacle to evolutionary modelling, because the MB torque cannot be predicted reliably across different physical regimes. Additional complications arise in such exotic cases as strongly magnetic CVs, where MB is significantly reduced because of the coupling of the donor and WD magnetospheres \citep{1994MNRAS.268...61L, 2020MNRAS.491.5717B}. To date, there is no universal MB recipe which would be able to predict the evolution of all systems consistently.

CVs can serve as useful laboratories for testing MB prescriptions. They are compact semi-detached binaries consisting of an accreting white dwarf (WD) and a Roche lobe filling companion star \citep{2003cvs..book.....W}. CVs are also X-ray sources, with typical luminosities spanning $L_X\sim10^{29}$--$10^{32}\,\mathrm{erg\,s^{-1}}$ (see, e.g. \citealt{2017PASP..129f2001M, 2024A&A...690A.374G}). The long-term evolution of CVs is governed by AML \citep{2023hxga.book..129B}, mainly through MB and gravitational radiation (GR) \citep{1967AcA....17..287P, 1984ApJS...54..443P}. Additional consequential angular momentum loss (CAML) is needed to account for the material leaving the system during nova eruptions \citep{1995ApJ...439..330K}. In the standard evolutionary picture of non-magnetic, hydrogen-rich CVs, MB dominates at longer orbital periods (see, e.g. \citealt{2011ApJS..194...28K}). At a period of $\sim3\,$hr, the donor star becomes fully convective and the strength of MB suddenly drops. This abrupt change in AML can initiate a detached phase, leaving the resulting WD+M-dwarf binaries effectively unobservable and giving rise to the so-called period gap \citep{1983A&A...124..267S}. The remaining GR shrinks the orbit until a period of $\sim 2\,$hr, when the donor fills its Roche lobe again and accretion restarts. At a period of $\sim80\,$min, the donor reaches a mass of $\sim0.08\,$M$_\odot$ and becomes a degenerate brown dwarf \citep{1999MNRAS.309.1034K}. The degenerate donor response to mass-transfer changes, and its radius now grows instead of shrinking, which leads to an increase in orbital period. The period gap at $\sim2-3\,$hr and period minimum at $\sim80\,$min are robust features of the CV period distribution, and can serve as key diagnostics for testing new stellar physics \citep{2024A&A...682L...7S}.

A recent development toward a more unified description of MB is the saturated, boosted and disrupted (SBD) MB prescription introduced by \citet{2024A&A...682A..33B}. It builds on the fact that MB saturates at faster rotation periods, which gained additional robust evidence in the last years from studies of different binaries \citep{2022MNRAS.517.4916E, 2025ApJ...995...19F}. Two additional multiplicative factors are introduced: one that boosts the braking strength above the fully convective boundary ($K$), and one that disrupts it once the donor becomes fully convective ($\eta$). While \citet{2024A&A...682A..33B} originally studied detached post common envelope WD+M-dwarf binaries (PCEBs), the SBD MB prescription has recently gained further support from studies of hot sdB subdwarfs plus main sequence (MS) binaries \citep{2024PASP..136l4201B} and CVs \citep{2025A&A...696A..92B}. Its apparent success across diverse binaries suggests that SBD MB may provide the basis for a more unified description of MB. However, the model remains essentially empirical, and the physical origin of the boost and disruption terms is still unclear. Previous SBD MB studies assumed constant values with $K\simeq\eta$ along the binary evolution, although the two factors likely reflect different underlying mechanisms.

Two possibilities appear particularly worth exploring. \citet{2024PASP..136l4201B} proposed that the boost may arise from irradiation-enhanced winds of the MS star, and \citet{2025A&A...696A..92B} suggested that the empirical treatment of the convective turnover time ($\tau_c$) may be a weak point of the model. On the one hand, the disrupted MB term $\eta$ should be connected to structural changes in the donor as it approaches the fully convective boundary, where $\tau_c$ is also expected to change abruptly \citep{2025ApJ...988..102G}. On the other hand, it is known that in close accreting binaries, accretion onto the compact object produces strong X-ray emission that can significantly heat the donor-facing side and induce a considerable stellar wind \citep{1973Ap&SS..23..117B, 1974A&A....31..249B, 1997ApJ...486..955I}. Such an irradiation-driven outflow may add significantly to the donor's intrinsic wind, thereby enhancing the mass loss that powers ordinary MB. In this study, we investigate whether these two mechanisms can account for the boost and disruption of MB in CVs by developing a physically motivated model and testing it with binary evolution calculations.

During the final stage of preparing this manuscript, \citet{2026A&A...708L..11B} revised SBD MB in CVs using the saturated prescription of \citet{2015ApJ...799L..23M}. They recalculated the global convective turnover time directly from the stellar structure, revised the saturation threshold, and found that, in this formulation, smaller empirical boost and disruption factors are sufficient to reproduce the main CV properties. Our results support that the treatment of $\tau_c$ is crucial and do not contradict the conclusions of \citet{2026A&A...708L..11B}. At the same time, using the original SBD MB formulation, we show that donor-structure effects in $\tau_c$ can directly drive MB disruption in saturated prescriptions that retain an explicit $\tau_c$-dependence, while irradiation-driven winds provide a plausible source of the remaining boost.

This paper is organized as follows. Section~\ref{sec:background} reviews the empirical form of SBD MB and describes the simulation setup in MESA. Section~\ref{sec:SBDisrupt} describes the structure-based calculation of the convective turnover time $\tau_c$ and its implications for the disruption of MB. Section~\ref{sec:SBDirad} introduces our irradiation prescription and its coupling to wind-driven enhancement of MB. Section~\ref{sec:itauSBD} presents the evolutionary results of the combined irradiation-coupled, structure-based convective turnover time i$\tau$SBD MB model. The implications of our results are discussed in Section~\ref{sec:discussion}, and we summarize our main conclusions in Section~\ref{sec:conclusions}.

\section{Background and simulation setup}
\label{sec:background}

\subsection{Saturated, boosted and disrupted magnetic braking}

Observations of various indicators of magnetic activity in Sun-like and low-mass stars reveal that activity increases with faster rotation but eventually saturates at a critical stage \citep{1984ApJ...279..763N, 2003A&A...397..147P, 2009ApJ...692..538R}. A crucial parameter governing this behaviour is the Rossby number $Ro$, defined as

\begin{equation}
    Ro=\frac{P_\mathrm{rot}}{\tau_c},
\end{equation}

\noindent where $P_\mathrm{rot}$ is the star's rotation period and $\tau_c$ is the convective turnover timescale. Empirically, magnetic activity correlates more tightly with $Ro$ than with rotation alone \citep{1984ApJ...279..763N}, indicating that the interplay between rotation and convection regulates stellar dynamos (see, e.g. Figure 1 from \citealt{2025ApJ...988..102G}). The transition between the unsaturated and saturated regimes occurs near a critical Rossby number $Ro_\mathrm{sat}\approx0.1$. By comparing to the solar Rossby number, this can be expressed through a dimensionless scaling parameter $\chi=Ro_\odot/Ro_\mathrm{sat}$ \citep{2015ApJ...799L..23M}, which allows the corresponding saturation period to be expressed as

\begin{equation}
\label{eq:p_sat}
    P_{\mathrm{sat}} = \frac{P_{\mathrm{rot},\odot}}{\chi}\frac{\tau_c}{\tau_{c,\odot}}.
\end{equation}

\noindent For our simulations, we adopted $\chi = 10$, $P_{\mathrm{rot,}\odot}=25\,$d and $\tau_{c,\odot}=15\,$d, which are close to the values used by \citet{2025A&A...696A..92B}. The convective turnover time $\tau_c$ increases toward lower masses \citep{2011ApJ...743...48W, 2018MNRAS.479.2351W}. For a relatively massive donor of $M_2=1.2\,M_\odot$, equation~\ref{eq:p_sat} yields $P_{\mathrm{sat}}\approx 1.8\,$d. This implies that typical CV donors with orbital periods shorter than one day are always expected to fall in the saturated regime.

The empirical saturation of stellar magnetic activity motivated the development of saturated MB prescriptions (e.g. \citealt{1988ApJ...333..236K, 2000ApJ...534..335S, 2003ApJ...582..358A}), in which the AML rate flattens for stars rotating faster than $P_{\mathrm{sat}}$. SBD MB builds on the saturated MB prescription by \citet{1995ApJ...441..865C}:

\begin{equation}
\label{eq:j_sat}
    \dot{J}_{\mathrm{MB, SAT}} = -C \left( \dfrac{R}{R_\odot} \right)^{1/2}
          \left( \dfrac{M}{M_\odot} \right)^{-1/2}
    \begin{cases}
          \left( \dfrac{P_{\mathrm{rot}}}{1\,\text{d}} \right)^{-3},
          & P_{\mathrm{rot}} \ge P_{\mathrm{sat}} \\[1.2em]
          \left( \dfrac{P_{\mathrm{sat}}}{1\,\text{d}} \right)^{-2}
          \left( \dfrac{P_{\mathrm{rot}}}{1\,\text{d}} \right)^{-1},
          & P_{\mathrm{rot}} < P_{\mathrm{sat}} 
    \end{cases}
\end{equation}

\noindent Here $C = 1.04\times10^{35}\,\mathrm{erg}$ is a calibrated constant (e.g. \citealt{2022MNRAS.517.4916E}), $M$ and $R$ are the star's mass and radius, $P_{\mathrm{rot}}$ is the star's rotation period and $P_{\mathrm{sat}}$ is the critical saturation period.

\citet{2024A&A...682A..33B} introduced two multiplicative factors to scale the saturated MB. These are the boosting factor $K$, which enhances MB, and the disruption factor $\eta$, which weakens MB once the star becomes fully convective:

\begin{equation}
    \dot{J}_{\mathrm{MB}} =
    \begin{cases}
        K\cdot\dot{J}_{\mathrm{MB, SAT}},\ \text{(rad. core + conv. envelope)} \\
        (K\cdot\dot{J}_{\mathrm{MB, SAT}})/\eta,\ \text{(fully convective).}
    \end{cases}
\end{equation}

\noindent \citet{2024A&A...682A..33B} showed that adopting $K\simeq\eta\gtrsim50$ in this framework reproduces the rise in the fraction of close systems among WD+M-dwarf binaries at the fully convective boundary. \citet{2024PASP..136l4201B} found evidence for SBD MB from the mass distribution of hot sdB+M-dwarf binaries, favouring $K\simeq\eta\gtrsim100$. \citet{2025A&A...696A..92B} applied SBD MB to CV evolution in MESA and found that $K\simeq\eta\simeq30\text{--}50$ provides a good match to key observational features. More recently, \citet{2026A&A...708L..11B} revised SBD MB for CVs using the saturated prescription of \citet{2015ApJ...799L..23M}, recalculated the global convective turnover times directly from the stellar structure, revised the saturation threshold, and found that smaller empirical boost and disruption factors, $K\sim20$ and $\eta\sim2\text{--}3$, are sufficient to reproduce the main CV properties.

\subsection{CV modelling in MESA}

We used the Modules for Experiments in Stellar Astrophysics (MESA) code (\citealt{2011ApJS..192....3P, 2013ApJS..208....4P, 2015ApJS..220...15P, 2018ApJS..234...34P, 2019ApJS..243...10P}; \citealt{2023ApJS..265...15J}, version r24.08.1)\footnote{The complete MESA setup used in this work is publicly available on Zenodo.}. Unless stated otherwise, we adopted $0.8\,M_\odot$ for both the initial WD and donor masses and set the initial orbital period to $P_{\mathrm{orb}}=1\,$d. We assumed the donor is tidally synchronised throughout the evolution. We did not evolve the WD and treated it as a point mass in the simulation. We assumed that all accreted material is expelled during nova eruptions, so the WD mass remains constant throughout the evolution \citep{2005ApJ...623..398Y}. The initial donor metallicity was set at the solar value of $Z=0.02$, and the simulation was terminated when the donor reached a mass of $0.05\,M_\odot$. We adopted a convection mixing-length parameter $\alpha_\mathrm{MLT}=1.82$, corresponding to the solar-calibrated value reported by \citet{2018ApJ...856...10J} for a grey atmosphere. The accretion rate $\dot{M}_\mathrm{acc}$ was calculated following the explicit Ritter scheme \citep{1988A&A...202...93R}: 

\begin{equation}
\label{eq:Ritter_mdot}
    \dot{M}_\mathrm{acc}\equiv-\dot{M}_2=\dot{M}_0\,\exp{\left(-\frac{R_\mathrm{2,R}-R_2}{H_P}\right)},
\end{equation}

\noindent where $M_2$ and $R_2$ are the donor's mass and radius, $R_\mathrm{2,R}$ is the radius of the donor's Roche lobe (computed using the Eggleton approximation; \citealt{1983ApJ...268..368E}), $H_P$ is the pressure scale height evaluated at the donor photosphere in the Roche potential, and $\dot{M}_0$ is a slowly varying function of the system parameters. AML via GR was included using the standard formula for a circular orbit (e.g. \citealt{1975ctf..book.....L}):

\begin{equation}
\label{eq:jdot_gr}
    \dot{J}_{\rm GR} = -\frac{32G^{7/2}M_1^{2} M_2^{2}\,\sqrt{M_1+M_2}}{5c^5a^{7/2}},
\end{equation}

\noindent where $M_1$ is the WD mass, $a$ is the orbital separation, $G$ is Newton's gravitational constant, and $c$ is the speed of light. For the overall effect of AML associated with mass expelled from the system, we adopted the empirical consequential angular momentum loss (eCAML; \citealt{2016MNRAS.455L..16S}):

\begin{equation}
\label{eq:eCAML}
    \frac{\dot{J}_{\mathrm{CAML}}}{J} = \frac{\nu}{M_1}\frac{\dot{M}_2}{M_2},
\end{equation}

\noindent where $J$ is the total orbital angular momentum and $\nu = 0.35\,M_\odot$. This prescription is now commonly adopted in CV evolution studies because it offers a practical solution to the WD mass problem (e.g. \citealt{2018MNRAS.478.5626B}), although its underlying physical mechanism remains unknown \citep{2020AdSpR..66.1080Z, 2024ApJ...977...34T}.

\section{Convective turnover time from stellar structure}
\label{sec:SBDisrupt}

\subsection{Impact on MB torque and CV evolution}

\begin{figure*}[h!]
    \centering
    \includegraphics[width = 0.49\linewidth]{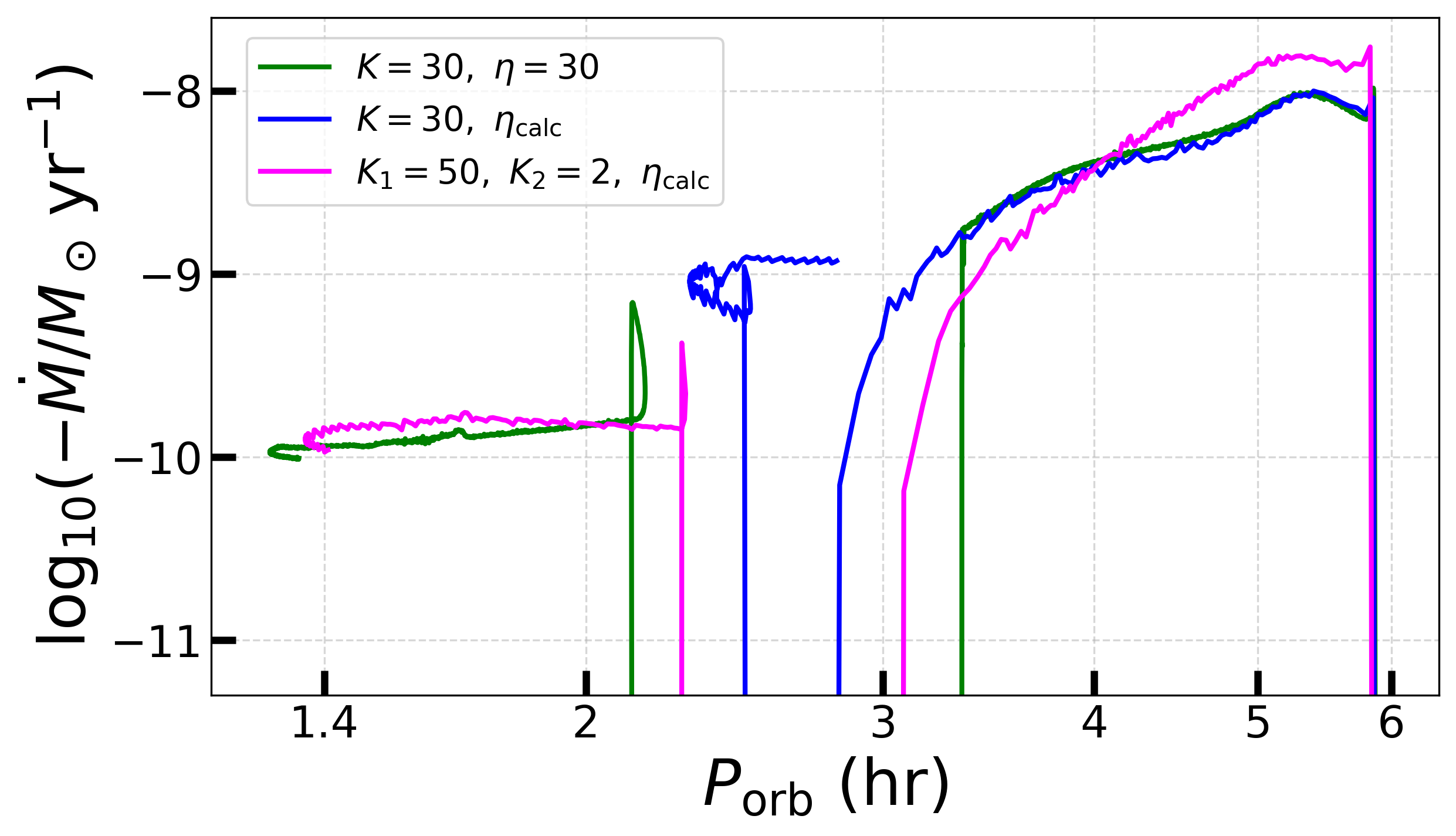}
    \includegraphics[width = 0.49\linewidth]{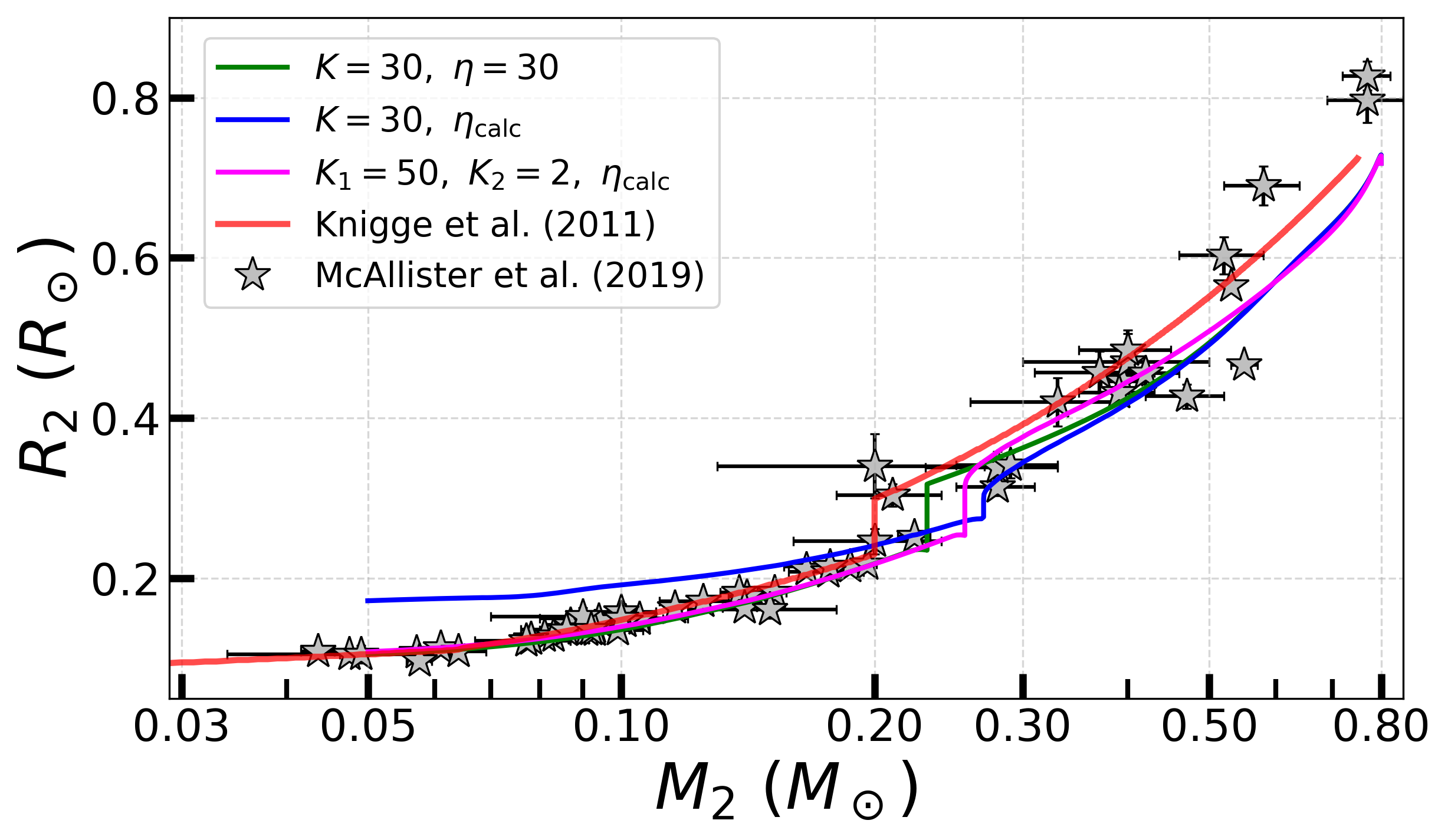}
    \includegraphics[width = 0.49\linewidth]{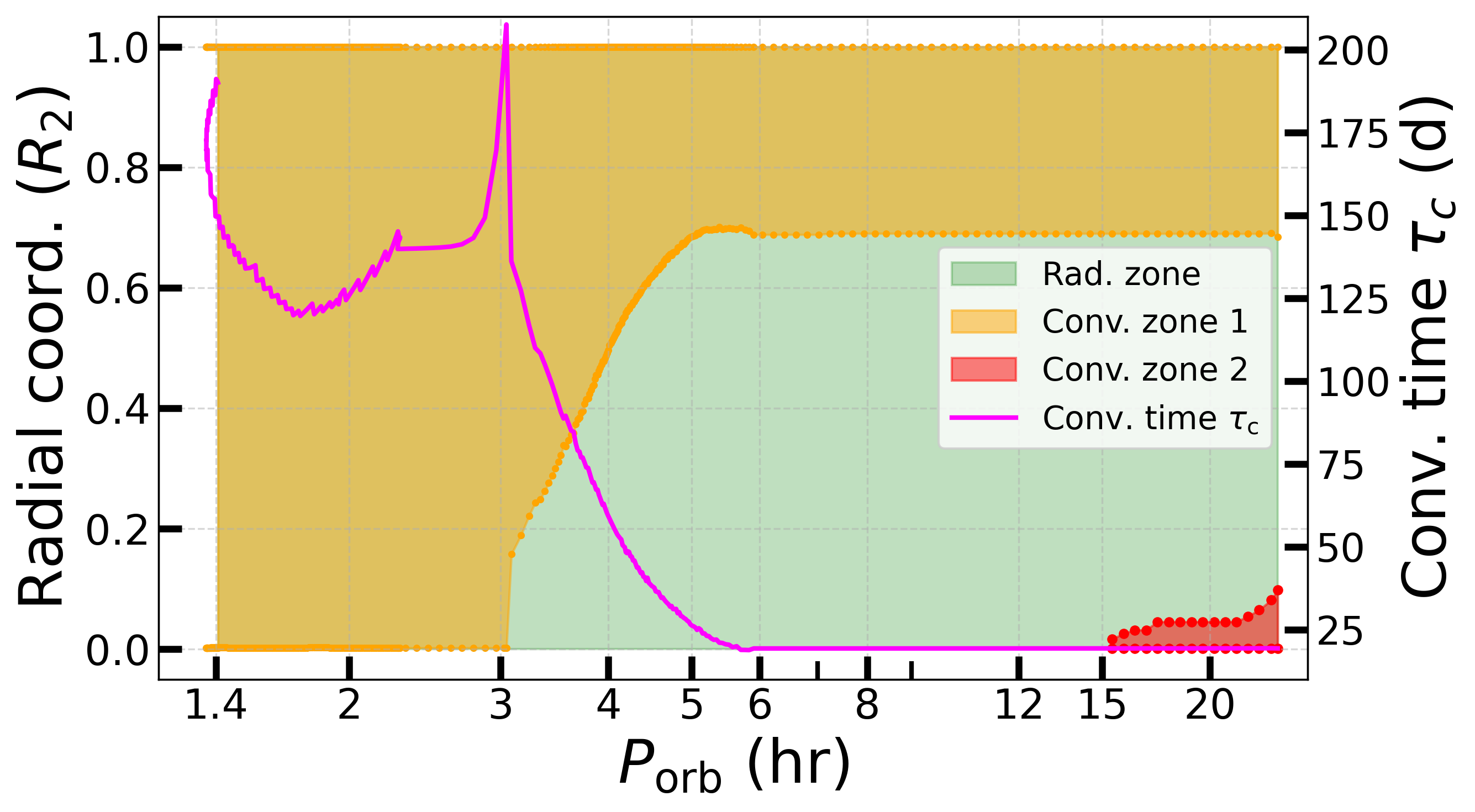}
    \includegraphics[width = 0.46\linewidth]{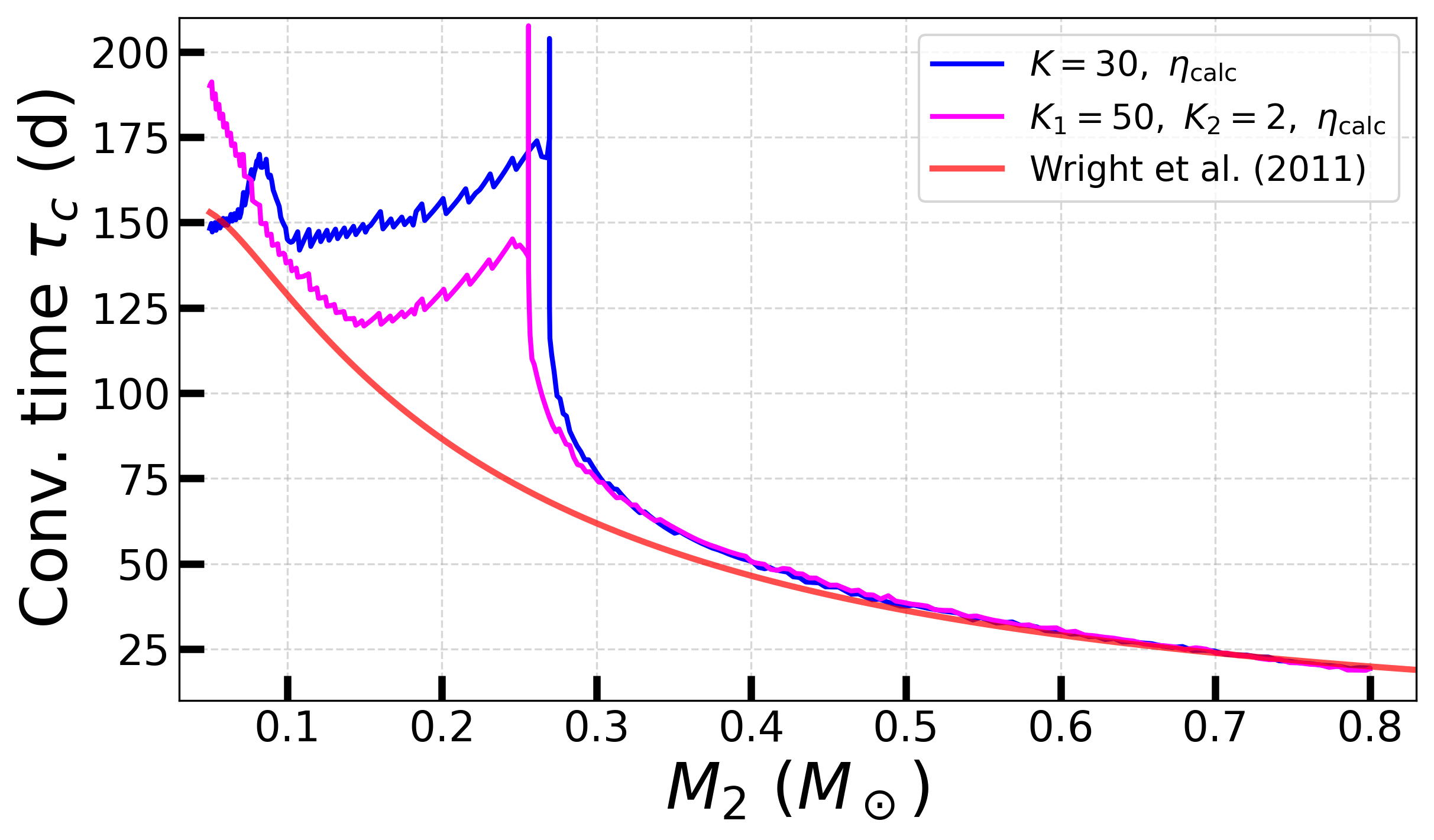}    
    \caption{CV evolution models using the SBD MB prescription with the convective turnover time $\tau_c$ computed directly from the donor structure. Upper left: mass transfer rate $\dot{M}$ versus orbital period $P_\mathrm{orb}$. Upper right: donor radius $R_2$ versus donor mass $M_2$. Grey star symbols show observational determinations from \citet{2019MNRAS.486.5535M}, and the red line is the semi-empirical donor sequence of \citet{2011ApJS..194...28K}. For comparison, the empirical SBD MB model with $K=\eta=30$ \citep{2025A&A...696A..92B} is shown in green. The blue track adopts the same constant boost, $K=30$. The magenta track uses a two-stage boost, transitioning from $K_1=50$ above the gap to $K_2=2$ once the system detaches and enters the gap. Lower left: evolution of the convective/radiative regions and the resulting $\tau_c(P_\mathrm{orb})$ for the two-stage model. Lower right: the $\tau_c$--$M_2$ relation for the two simulations using new $\tau_c$ calculation compared with the empirical $\tau_c(M)$ relation of \citet{2011ApJ...743...48W}. A pronounced spike in $\tau_c$ appears when the donor becomes fully convective.}
    \label{fig:SBDisrupt}
\end{figure*}

Equations \eqref{eq:p_sat} and \eqref{eq:j_sat} show that the convective turnover time, $\tau_c$, is a key parameter for MB in the saturated regime, with $\dot{J}_{\mathrm{MB, SAT}}\propto\tau_c^{-2}$. Previous SBD MB studies have adopted the smooth, mass-only empirical relation $\tau_c(M)$ from \citet{2011ApJ...743...48W}. This prescription is potentially unreliable for fully convective low-mass M dwarfs, where observational constraints are sparse \citep{2018MNRAS.479.2351W, 2020A&A...638A..20M}. Moreover, empirical $\tau_c$ relations derived from single stars may not apply directly to mass-losing donors in CVs. Recently, \citet{2025ApJ...988..102G} found, both empirically and theoretically, a sharp increase in $\tau_c$ near $0.35\,M_\odot$, where typical M-dwarfs become fully convective. This motivates the idea that the MB torque may change abruptly at the fully convective boundary. For example, a $\sim5\times$ increase in $\tau_c$ is sufficient to produce the desired disruption factor $\eta\sim25$.

We followed the approach of \citet{2025ApJ...988..102G}\footnote{\url{https://zenodo.org/records/15680676}} in the calculation of $\tau_c$ from the stellar structure using MESA. Physically, $\tau_c$ characterizes the timescale of convective motions in the stellar interior. The size and velocity of convective cells vary with depth, and within mixing-length theory $\tau_c$ can be defined locally as (e.g. \citealt{2005PhR...417....1B}):

\begin{equation}
    \tau_c(r) = \frac{H_{\mathrm{P}}(r)}{v_c(r)},
\end{equation}

\noindent where $H_{\mathrm{P}}(r)=P(r)/g(r)\rho(r)$ is the local pressure scale height, and $v_c(r)$ is the local convective velocity. Magnetic dynamos are believed to operate near the base of the convective envelope, at the interface with the radiative core, in a thin shear layer known as the tachocline \citep{1955ApJ...122..293P, 1975ApJ...198..205P, 1992A&A...265..106S}. This motivates evaluating $\tau_c$ near the base of the convective envelope $r_{\mathrm{BCE}}$. However, as the star becomes fully convective, $r_{\mathrm{BCE}}$ matches its centre and divergence would occur as $g(r_{\mathrm{BCE}}\to0)\to0$ and $v_c(r_{\mathrm{BCE}}\to0)\to0$. A practical solution is to evaluate $\tau_c$ at some distance from the bottom of the convective envelope. For example, \citet{1985ApJ...299..286G} proposed using one scale height above ($r=r_{\mathrm{BCE}}+H_{\mathrm{P}}(r_{\mathrm{BCE}})$), and \citet{2025ApJ...988..102G} adopted half a local scale height above ($r=r_{\mathrm{BCE}}+0.5H_{\mathrm{P}}(r_{\mathrm{BCE}})$). We decided to use one scale height above as it reproduces the empirical relation of \citet{2011ApJ...743...48W} at $M\gtrsim0.5\,M_\odot$.

The results of incorporating the new structure-based $\tau_c$ into the SBD MB model for CV evolution are shown in Fig.~\ref{fig:SBDisrupt}. For comparison, we also show the empirical SBD MB with $K=\eta=30$ \citep{2025A&A...696A..92B}. Upper panels show the mass transfer rate $\dot{M}$ versus orbital period $P_\mathrm{orb}$ and donor radius $R_2$ versus mass $M_2$. The lower right panel shows the $\tau_c$--$M_2$ relation for the simulations using new $\tau_c$ calculation compared with the empirical $\tau_c(M)$ relation of \citet{2011ApJ...743...48W} used in previous studies. For the models with new $\tau_c$ calculation, $\eta$ is calculated simply as $\eta=(\tau_c^\mathrm{calc}/\tau_c^\mathrm{emp})^2$, where $\tau_c^\mathrm{emp}$ is given by the empirical relation of \citet{2011ApJ...743...48W}. The structure-based $\tau_c$ broadly agree with the empirical relation at $M_2\gtrsim0.5\,M_\odot$. In typical M-dwarfs the fully convective boundary corresponds to $M\approx0.35\,M_\odot$ \citep{1997A&A...327.1039C}. In the case of CV donors, the fully convective boundary is shifted to lower masses $M_2\approx0.2\text{--}0.25\,M_\odot$. Our simulations show a spike in $\tau_c$ at $M_2\approx0.25\,M_\odot$ corresponding to a $\sim2-3\times$ increase. This spike directly triggers a quick MB disruption by a factor of $\eta\sim5$, and the system naturally detaches entering the period gap. Because the $\tau_c$ values are systematically larger than the empirical ones for $M_2\lesssim0.5\,M_\odot$, we found that a stronger boosting factor $K=50$ is needed above the gap. However, the system then re-establishes contact too early and the post-gap accretion rates are too high, indicating that the effective boost must drop once detachment occurs. We therefore reduced the boost when the system detaches (for example when $\dot{M}\lesssim10^{-20}\,M_\odot\,\mathrm{yr}^{-1}$) and found that a residual value $K_2=2$ works well. The lower left panel of Fig.~\ref{fig:SBDisrupt} shows the evolution of the convective/radiative regions and the resulting $\tau_c(P_\mathrm{orb})$ for the two-stage $K_1\rightarrow K_2$ model. As the radiative core shrinks and the convective envelope deepens toward full convection, the locally-evaluated $\tau_c$ increases. 

We note that the present model involves some circularity. By assuming that the relevant dynamo operates near the interface between the convective envelope and the radiative core, the weakening of MB is naturally linked to the disappearance of that interface as the donor approaches full convection. In this sense, the MB disruption is built into the adopted physical picture. At present, the precise role of the tachocline in dynamos of low-mass stars, and the relation between partially and fully convective dynamos remain uncertain \citep{2016ApJ...819..104G, 2018MNRAS.479.2351W, 2023SSRv..219...58K, 2024NatAs...8..223L}. This leaves room for alternative assumptions about the dynamo-relevant region and the most appropriate definition of $\tau_c$.

\subsection{Dependence on the $\tau_c$ definition}

To assess how strongly our results depend on the adopted definition of $\tau_c$, we repeated the simulations using different prescriptions for $\tau_c$. In addition to our adopted local prescription evaluated at $r=r_{\mathrm{BCE}}+H_{\mathrm{P}}(r_{\mathrm{BCE}})$, we considered the alternative local prescription of \citet{2025ApJ...988..102G}, evaluated at $r=r_{\mathrm{BCE}}+0.5H_{\mathrm{P}}(r_{\mathrm{BCE}})$, as well as the global envelope-integrated definition:

\begin{equation}
\tau_c = \int_{R_{\mathrm{BCE}}}^{R_*} \frac{dr}{v_{\mathrm{conv}}(r)}.
\end{equation}

\noindent This definition has, for example, been adopted in the CARB MB prescription of \citet{2019ApJ...886L..31V}\footnote{\url{https://zenodo.org/records/3647683}} and in the recent recalibration of SBD MB by \citet{2026A&A...708L..11B}\footnote{\url{https://zenodo.org/records/19010875}}. Unlike the local prescriptions, the global definition does not assume the tachocline to be the unique dynamo-relevant location. Therefore, it has the conceptual advantage that the increase in $\tau_c$ emerges from the structure of the entire convective envelope.

\begin{figure}[h!]
    \centering
    \includegraphics[width = 1.0\linewidth]{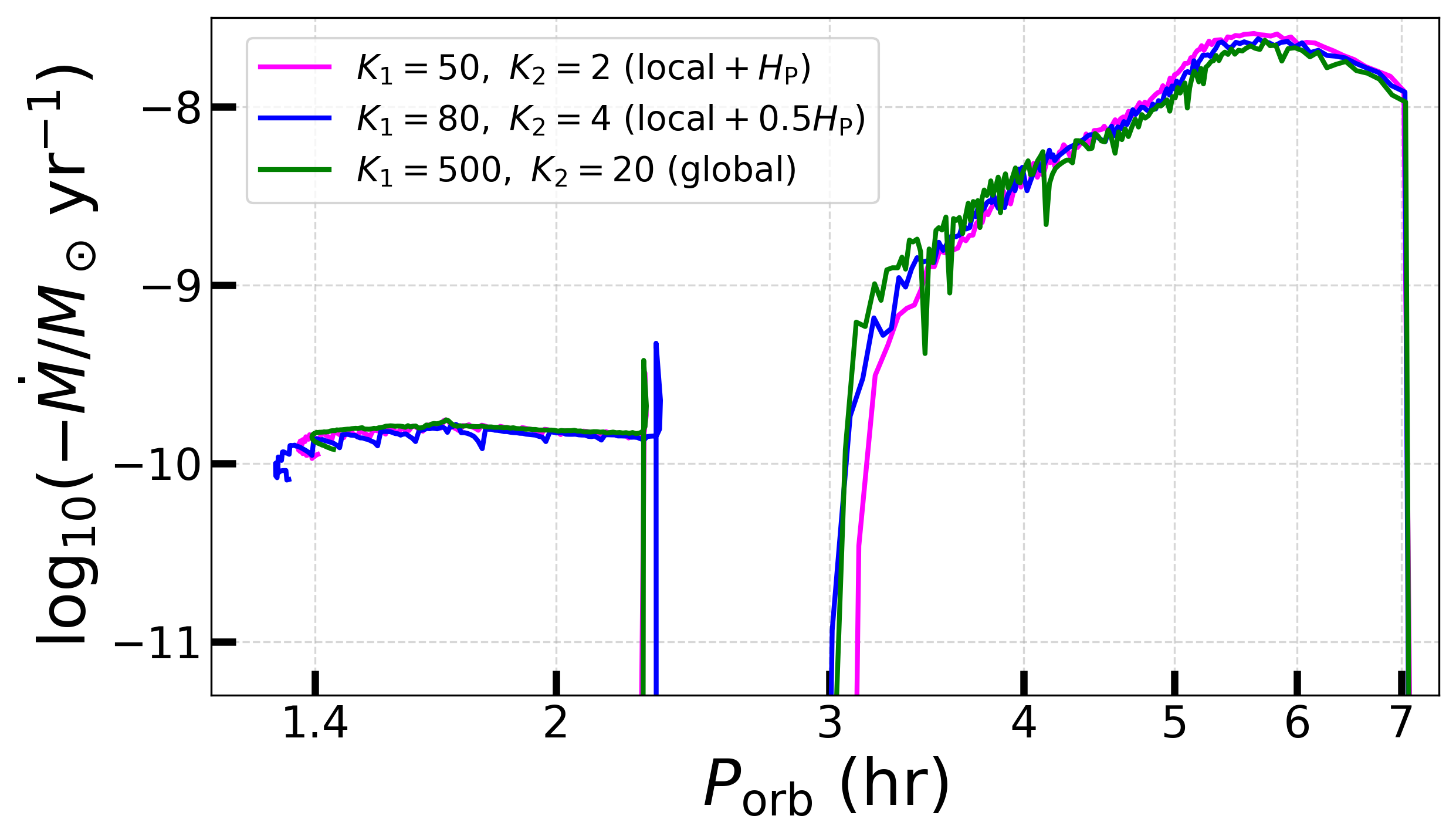}
    \includegraphics[width = 1.0\linewidth]{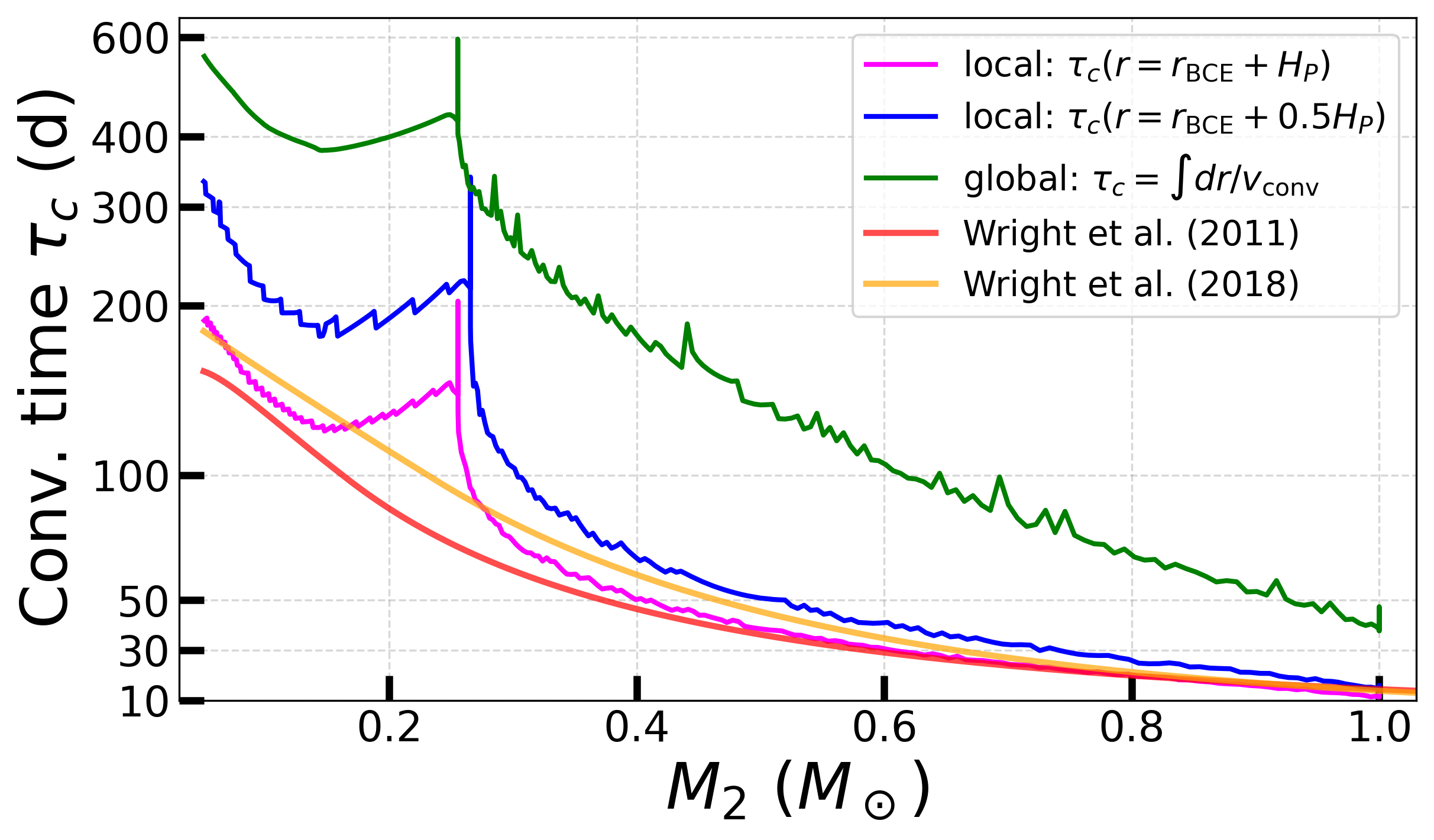}
    \caption{Sensitivity of the CV evolution and the $\tau_c$--$M_2$ relation to the adopted convective turnover time prescription (see text). For each case, the boost factors $K_1$ and $K_2$ were adjusted to preserve broadly consistent CV evolutionary tracks. The corresponding $\tau_c$--$M_2$ relations are compared with the empirical relations of \citet{2011ApJ...743...48W} and \citet{2018MNRAS.479.2351W}. All three prescriptions produce a pronounced increase in $\tau_c$ near the fully convective boundary with similar amplitudes, although normalisation depends on the adopted definition.}
    \label{fig:SBDisrupt_def}
\end{figure}

Figure~\ref{fig:SBDisrupt_def} compares CV evolutionary tracks computed with these three $\tau_c$ prescriptions. For each case, we allowed the MB boost parameters to be adjusted so as to preserve a broadly consistent CV evolutionary picture, including the period gap and period minimum. To illustrate the behaviour over a wider mass range, we used an initial donor mass of $1\,M_\odot$. The figure also compares the resulting $\tau_c$--$M_2$ relations with the empirical relation of \citet{2011ApJ...743...48W} and the updated relation from \citet{2018MNRAS.479.2351W}. Most importantly, all three prescriptions show a pronounced increase in $\tau_c$ as the donor approaches the fully convective boundary at $M_2 \approx 0.25\,M_\odot$. The relative amplitude of this feature is very similar in all cases, corresponding to a $\sim 2$--$3\times$ increase compared to the values immediately above the transition. The main difference between the prescriptions lies instead in the overall normalisation of $\tau_c$. The local prescription evaluated at $r=r_{\rm BCE}+H_P$ remains close to the empirical $\tau_c$--$M$ relation of \citet{2011ApJ...743...48W}, while the alternative local prescription at $r=r_{\rm BCE}+0.5H_P$ is more consistent with the updated relation from \citet{2018MNRAS.479.2351W}. By contrast, the global envelope-integrated definition yields systematically larger values of $\tau_c$, likely reflecting the different averaging procedure rather than the same physical timescale being measured. As a result, in our calibration the global definition requires MB boost factors roughly an order of magnitude larger than the adopted local prescription evaluated at $r=r_{\rm BCE}+H_P$ in order to reproduce similar CV evolutionary tracks.

Taken together, this analysis indicates that the initial MB disruption by a factor of $\eta \sim 5$ is robust across the different $\tau_c$ prescriptions explored here, and is primarily driven by the spike in $\tau_c$. The CV period gap is initiated by the donor's structural response as it approaches full convection. The subsequent reduction of the boost below the gap suggests that the boosting mechanism should also depend on the donor's structure. In the remainder of this paper, we adopt the local definition evaluated at $r=r_{\rm BCE}+H_P$, as it allows direct comparison with previous SBD MB studies.

\section{Donor irradiation by the accreting WD}
\label{sec:SBDirad}

We used a simplified model to assess whether irradiation-driven winds in CVs can produce a significant enhancement of MB and thereby account for the empirical boost factor $K$. We note that irradiation is an inherently anisotropic process that cannot be modelled explicitly in a 1D stellar structure code. The solution is to approximate it as an effective, spherically averaged heat source deposited in the donor's outer envelope, which is appropriate for studying secular evolution.

\subsection{The irradiation model}

We computed the X-ray luminosity assuming that half of the gravitational energy of the accreted matter is released in the hot boundary layer, which generates X-rays (see, e.g. \citealt{1985ApJ...292..535P, 2025PASP..137a4201R}):

\begin{equation}
\label{eq:Lx}
    L_{\mathrm{X}} = \alpha_{\mathrm{acc}}\frac{1}{2}\frac{G M_{\mathrm{WD}} \dot{M}_{\mathrm{acc}}}{2R_{\mathrm{WD}}},
\end{equation}

\noindent where $\alpha_{\mathrm{acc}}$ is the radiative efficiency of accretion, $M_{\mathrm{WD}}$ and $R_{\mathrm{WD}}$ are the WD mass and radius, and $\dot{M}_{\mathrm{acc}}$ is the mass accretion rate. The additional factor of $1/2$ reflects the assumption that only half of the boundary-layer X-ray emission escapes outward, while the other half is intercepted and absorbed by the WD. For a given WD mass, we computed $R_{\mathrm{WD}}$ from the zero-temperature WD mass–radius relation of \citet{1972ApJ...175..417N}. The accretion efficiency $\alpha_{\mathrm{acc}}$ is an unknown parameter, which we try to relate to realistic values from observations. \citet{2025PASP..137a4201R} obtained that $\alpha_{\mathrm{acc}}=0.02\text{--}0.3$ can adequately reproduce the observed X-ray luminosities of the CV population. 

To compute the amount of X-ray luminosity $L_{\mathrm{X}}$ absorbed by the donor, we adopted the point–source irradiation model \citep{1996ApJ...467..761K, 2000A&A...360..969R, 2004A&A...423..281B}. In this model, the accreting WD is treated as an isotropic point source at separation $a$. The local irradiating flux on a surface element depends on its orientation and position on the donor star. For a surface element at colatitude $\theta$, the irradiating flux normal to the surface can be written as:

\begin{equation}
\label{eq:F_irad}
    F_\mathrm{X, irr}(\theta) = \alpha_{\mathrm{irr}}\frac{L_\mathrm{X}}{4\pi a^2}h(\theta),
\end{equation}

\noindent where $\alpha_{\mathrm{irr}}$ is an irradiation efficiency that accounts for disc shadowing and the donor's albedo, and $h(\theta)$ is the geometrical factor (see, e.g. \citealt{2004A&A...423..281B} for an illustration):

\begin{equation}
\label{eq:h(theta)}
    h(\theta) = \frac{\cos(\theta)-\cos(\theta_\mathrm{max})}{(1-2\cos(\theta)\cos(\theta_\mathrm{max})+\cos^2(\theta_\mathrm{max}))^{3/2}},
\end{equation}

\noindent where $\theta_\mathrm{max}=\arccos(R_2/a)$ is the maximum colatitude for direct irradiation. Integrating $F_\mathrm{X, irr}(\theta)$ over the illuminated (day-side) surface of the donor yields the total power absorbed by the donor:

\begin{equation}
\label{eq:Pabs_int}
    P_\mathrm{abs} = \int_\mathrm{day} F_\mathrm{X, irr}(\theta) \mathrm{d}A = \alpha_{\mathrm{irr}}\frac{L_\mathrm{X}}{4\pi a^2}\int_0^{2\pi}\int_0^{\theta_\mathrm{max}}h(\theta)R_2^2 \sin(\theta)\,\mathrm{d}\phi\, \mathrm{d}\theta.
\end{equation}

\noindent The integral can be evaluated to give the final formula accounting for the geometry:

\begin{equation}
\label{eq:Pabs}
    P_\mathrm{abs} = \alpha_{\mathrm{irr}}\frac{L_\mathrm{X}R_2^2}{4\pi a^2}f_\mathrm{geom},
\end{equation}

\noindent with the geometrical factor given by:

\begin{equation}
\label{eq:f_geom}
    f_\mathrm{geom} = 2\pi\frac{1-\sqrt{1-(R_2/a)^2}}{(R_2/a)^2}.
\end{equation}

\noindent In the limit $R_2/a\ll1$, one recovers $f_\mathrm{geom}\to\pi$, and $P_\mathrm{abs}$ reduces to the familiar projected–area expression $P_\mathrm{abs} = \alpha_{\mathrm{irr}} L_\mathrm{X}(R_2^2/4 a^2)$. For typical CV parameters, we found that the geometrical factor induces only a modest $<10\%$ difference compared to the simple projected-area expression. The irradiation efficiency $\alpha_{\mathrm{irr}}$ is our second unknown parameter, in addition to $\alpha_{\mathrm{acc}}$. In the idealised case where the accretion disc lies exactly in the orbital plane and does not obscure the line between the WD and the donor, disc shadowing becomes negligible and the efficiency reduces to $\alpha_{\mathrm{irr}}\approx(1-A_\mathrm{X})$. Here, $A_\mathrm{X}$ is the X-ray albedo of the donor describing how much of the incident X-ray flux is reflected. Several studies of X-ray illuminated stellar atmospheres show that at most $\sim30-50\%$ of the X-ray flux could be reflected \citep{1974A&A....31..249B, 1976A&A....46..189F, 1981ApJ...243..970L}. Therefore, the irradiation efficiency should be about $\alpha_{\mathrm{irr}}\gtrsim0.5$.

We deposited $P_\mathrm{abs}$ as an additional, spherically averaged heat source in the donor's outer envelope. We parametrised the deposition depth by the column depth $\Sigma$:

\begin{equation}
\Sigma(r) = \int_r^{R_2} \rho(r')\,\mathrm{d}r',
\end{equation}

\noindent which describes the mass per unit surface area above radius $r$. In discretised form, the column depth at the outer face of zone $k$ is obtained by summing inward from the surface:

\begin{equation}
\Sigma_k = \sum_{j=1}^{k}\frac{\Delta m_j}{4\pi r_j^2},
\end{equation}

\noindent where $\Delta m_j$ is the mass of zone $j$ and $r_j$ is the representative shell radius. By default, MESA deposits the additional heating uniformly down to a prescribed column depth (with the standard choice $\Sigma=1\,\text{g}\,\text{cm}^{-2}$). Instead, we adopted an exponential deposition law:

\begin{equation}
\label{eq:heat_exp}
\frac{dP}{d\Sigma}=\frac{P_\mathrm{abs}}{\Sigma_\mathrm{char}}
\exp\!\left(-\frac{\Sigma}{\Sigma_\mathrm{char}}\right),
\end{equation}

\noindent where $\Sigma_\mathrm{char}$ is the characteristic column depth that controls how concentrated the heating is toward the near-surface layers. In principle, a more realistic treatment would account for the incident X-ray spectrum and assign energy-dependent penetration depths \citep{2013AAS...22120205Q}\footnote{See also J.~Quintin's thesis for a detailed discussion on irradiation-depth: \url{https://jerome-quintin.github.io/assets/docs/JeromeQuintin_BSc_thesis.pdf}.}. A full spectral treatment is beyond the scope of this study and we adopted $\Sigma_\mathrm{char}=3\,\text{g}\,\text{cm}^{-2}$. We verified that varying this parameter over the range $\Sigma_\mathrm{char}=1$--$30\,\mathrm{g\,cm^{-2}}$ leaves the evolutionary tracks largely unchanged. For smaller values, the stronger concentration of heating in the outermost layers can occasionally trigger unstable mass transfer under strong irradiation. For a zone bounded by $\Sigma_{k-1}$ and $\Sigma_k$, the deposited power is:

\begin{align}
\Delta P_k
&= P_{\rm abs}\left[\exp\!\left(-\frac{\Sigma_{k-1}}{\Sigma_{\rm char}}\right)
-\exp\!\left(-\frac{\Sigma_{k}}{\Sigma_{\rm char}}\right)\right] \\
&= P_{\rm abs}\,\exp\!\left(-\frac{\Sigma_{k-1}}{\Sigma_{\rm char}}\right)
\left[1-\exp\!\left(-\frac{\Delta\Sigma_k}{\Sigma_{\rm char}}\right)\right],
\end{align}

\noindent which guarantees the correct normalisation $\sum_k \Delta P_k \simeq P_{\rm abs}$. We converted $\Delta P_k$ to a specific heating rate:

\begin{equation}
\epsilon_k = \frac{\Delta P_k}{\Delta m_k},
\end{equation}

\noindent and injected it into the stellar energy equation via MESA's per-cell extra-heating term (\texttt{extra\_heat}). We prefer the exponential profile as given by Eq.~\ref{eq:heat_exp} because irradiation is expected to be absorbed predominantly in the upper layers, and a uniform-cutoff prescription introduces a sharp discontinuity at the base of the heated region that can lead to numerical instabilities.

We considered an additional donor wind component that is induced by irradiation heating. The motivation is that if the energy deposited by irradiation in the donor's outer layers is sufficient to raise the local temperature and drive expansion, an extra outflow may be launched from the irradiated hemisphere. We treated this process in a highly simplified, energy-limited way and did not attempt to model the full hydrodynamics of wind generation. In general, the wind dynamics depend on hydrodynamical properties of the heating and cooling regimes, as well as the incident X-ray spectrum of irradiation \citep{1977ApJ...215..276B, 1989ApJ...343..292R, 1993ApJ...410..281T}. To lift material from the stellar surface out of the donor's potential well, the minimum required specific energy is of order the escape energy $E_{\mathrm{esc}}\approx GM_2/R_2$. Since only a fraction of the irradiation-induced energy can be converted into mechanical work (with the rest radiated away), a wind efficiency coefficient $\alpha_\mathrm{wind}$ must be introduced. Comparing the available driving power $\alpha_{\mathrm{wind}} P_{\mathrm{abs}}$, with the mechanical power carried by the wind $\dot{M}_{\mathrm{wind,irr}}E_{\mathrm{esc}}$, we obtained:

\begin{equation}
\label{eq:Mdot_wind}
    \dot{M}_\mathrm{wind,irr} = \alpha_{\mathrm{wind}}\frac{P_\mathrm{abs}R_2}{GM_2}.
\end{equation}

\noindent It is believed that $\alpha_\mathrm{wind}$ is a relatively small parameter. \citet{1993ApJ...410..281T} estimated $\alpha_\mathrm{wind}\sim10^{-1}-10^{-3}$ in LMXBs from hydrodynamical calculations. \citet{2006MNRAS.366.1415J} and \citet{2016ApJ...830..131C} used $\alpha_\mathrm{wind} \sim 10^{-3}-10^{-5}$ to simulate the evolution of compact X-ray binaries with black holes and neutron stars. \citet{2019ApJ...887..201X} also used $\alpha_\mathrm{wind} \sim 10^{-2}-10^{-3}$ to explain the formation of the compact binary 2A 1822-371. We found that for typical CV parameters and plausible efficiencies, the irradiation-driven wind loss is typically of order $\dot{M}_\mathrm{wind,irr}\lesssim10^{-11}\,M_\odot\,\text{yr}^{-1}$. The impact of this extra mass loss on the mass-transfer budget is small compared to the accretion-driven mass-transfer rate but the irradiation-driven wind can still be a significant fractional enhancement relative to the donor's intrinsic (base) wind $\dot{M}_\mathrm{wind,base}$, and may therefore produce a substantial boost of the MB torque. To express this MB boosting, we compared the irradiation-induced wind to the base wind:

\begin{equation}
\label{eq:Kboost}
    K=1+ \left( \frac{\dot{M}_\mathrm{wind,\,irr}}{\dot{M}_\mathrm{wind,\,base}} \right)^{\beta},
\end{equation}

\noindent where $\beta$ describes how the MB torque scales with wind mass loss $\dot{J}_{\mathrm{MB}}\propto\dot{M}_\mathrm{wind}^\beta$\footnote{Here $K$ is defined as the dimensionless enhancement of the MB torque relative to the baseline torque associated with the donor's intrinsic wind,
$K \equiv \dot J_{\rm MB,tot}/\dot J_{\rm MB,base}$. The form adopted in Eq.~\ref{eq:Kboost} treats the irradiation-driven wind as a separate contribution to the MB torque, that is, $\dot J_{\rm MB,tot}=\dot J_{\rm MB,base}+\dot J_{\rm MB,irr}$. An alternative choice would be to define the total wind mass-loss rate, $\dot M_{\rm wind,tot}=\dot M_{\rm wind,base}+\dot M_{\rm wind,irr}$, which leads to $K=(1+\dot M_{\rm wind,irr}/\dot M_{\rm wind,base})^{\beta}$. We verified that this alternative mainly affects the low-boost regime ($K\lesssim3$), with no significant change in the results.}. In modern MB formulations the dependence of torque on other parameters is expressed through an exponent $m$, which relates to our parameter as $\beta=1-2m$ \citep{2015ApJ...799L..23M}. Magnetohydrodynamical simulations suggest that $m$ is determined primarily by the magnetic field geometry and wind acceleration profile, and likely $m\sim0.2-0.25$ \citep{2015ApJ...798..116R}. Therefore, $\beta$ is expected to be of order $\beta\sim0.5$.

Unfortunately, the base wind mass-loss rates of M-dwarfs $\dot{M}_{\mathrm{wind,\,base}}$ remain poorly constrained. \citet{2021ApJ...915...37W} report estimates for 17 M-dwarfs and find that almost all of them satisfy $\dot{M}_{\mathrm{wind,base}} \lesssim \dot{M}_\odot$ (see also \citealt{2023MNRAS.524.5096W}). Although there exist prescriptions that relate wind mass-loss rates to stellar parameters (e.g. \citealt{1975MSRSL...8..369R, 1988A&AS...72..259D, 2005ApJ...630L..73S, 2015A&A...577A..28J}), these relations are highly uncertain for low-mass M dwarfs, particularly near the fully convective boundary. In our case, we found that such prescriptions can also be inconsistent with empirical upper limits \citep{2021ApJ...915...37W, 2023MNRAS.524.5096W}. We therefore adopted the simplification of a constant base wind in all simulations:

\begin{equation}
\label{eq:mdot_value}
\dot{M}_{\mathrm{wind,\,base}} = 0.1\,\dot{M}_\odot,
\end{equation}

\noindent where for the solar wind mass-loss rate we used $\dot{M}_\odot = 1.4\times10^{-14}\,M_\odot\,\mathrm{yr}^{-1}$ \citep{2015A&A...577A..27J}. We note that this ad-hoc choice mainly affects the normalisation of the MB boost $K$. For alternative values of $\dot{M}_{\mathrm{wind,base}}$, the desired $K$ can still be obtained by varying the uncertain efficiency factors $\alpha_\mathrm{acc}, \alpha_\mathrm{irr}, \alpha_\mathrm{wind}$. In practice, the value adopted in Eq.~\ref{eq:mdot_value} allows these efficiencies to remain within the plausible bounds discussed above.

\subsection{Mass-transfer cycles}

\begin{figure}[h!]
    \centering
    \includegraphics[width = 1.0\linewidth]{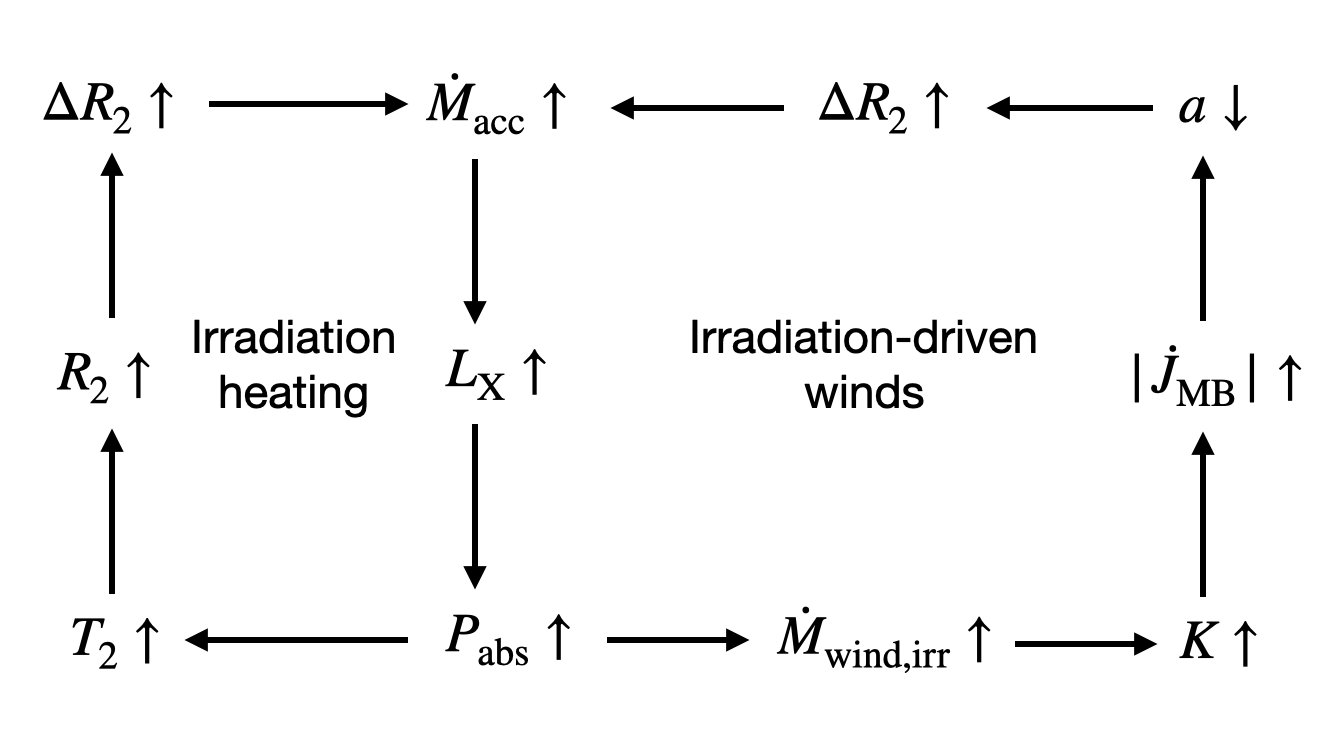}
    \caption{Schematic diagram of the two positive feedback loops in our irradiation model that make the mass-transfer response stiff and result in mass-transfer cycles. An increase in the accretion rate $\dot{M}_\mathrm{acc}$ raises the accretion luminosity $L_X$ and the absorbed irradiation power $P_\mathrm{abs}$. In the heating loop (left), irradiation modifies the donor's outer boundary conditions, driving expansion ($R_2\uparrow$), which increases the Roche lobe overfill $\Delta R_2 \equiv R_2 - R_{2,\mathrm{R}}$ and hence further enhances $\dot{M}_{\rm acc}$ (see Eq.~\ref{eq:Ritter_mdot}). In the wind loop (right), irradiation drives a wind $\dot{M}_\mathrm{wind,irr}$ that increases the MB boost factor $K$ and the MB torque $|\dot{J}_\mathrm{MB}|$, shrinking the orbit and Roche lobe ($R_{2,\mathrm{R}}\downarrow$), again increasing $\Delta R_2$ and $\dot{M}_\mathrm{acc}$.}

    \label{fig:irad_feedback}
\end{figure}

The irradiation prescription above couples the instantaneous mass-transfer rate $\dot{M}_{\mathrm{acc}}$ to the donor structure and to the MB torque. Because accretion rate responds extremely sensitively to small changes in the degree of overfill (see Eq.~\ref{eq:Ritter_mdot}), these couplings form a stiff feedback that can naturally produce relaxation oscillations (mass-transfer cycles). Irradiation-driven cycles have previously been found even when irradiation affects the donor only through heating/flux blocking, without any additional wind-driven MB enhancement (e.g. \citealt{2000A&A...360..969R,2004A&A...423..281B}). In our framework, the same structural feedback is present, but irradiation also drives an additional wind component that boosts the MB torque, introducing a second positive feedback loop. Figure~\ref{fig:irad_feedback} summarizes the two loops.

\begin{figure}[h!]
    \centering
    \includegraphics[width = 1.0\linewidth]{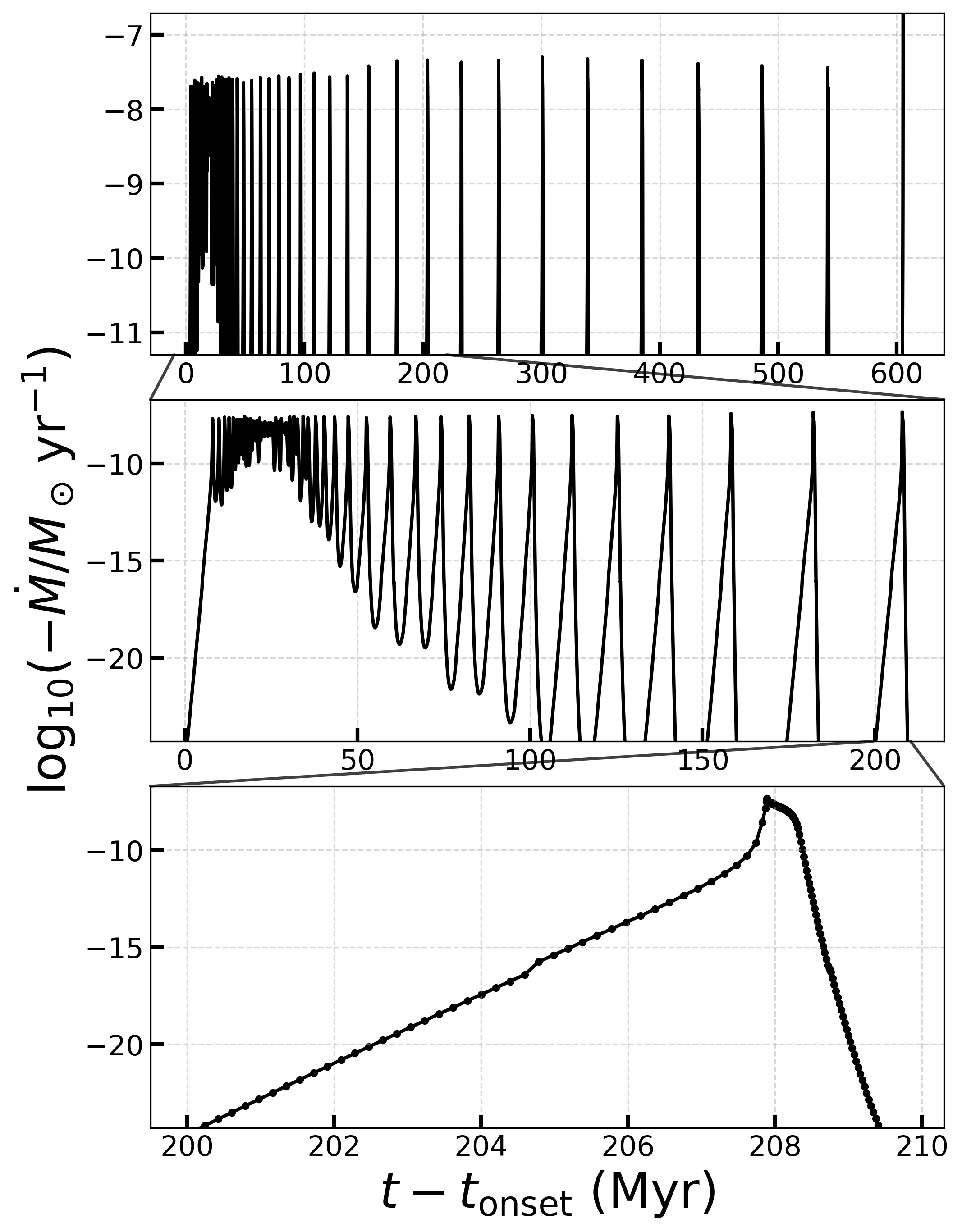}
    \caption{Mass-transfer rate as a function of time since the onset of mass transfer for the irradiated model (see text for details). The evolution exhibits recurrent mass-transfer cycles. After $\approx 600\,\mathrm{Myr}$, $\dot{M}$ rises to $\gtrsim 10^{-6}\,M_\odot\,\mathrm{yr^{-1}}$ and the simulation terminates. The middle and bottom panels show successive zoom-ins of the time intervals highlighted. The markers in the bottom panel indicate the discrete simulation steps, illustrating the temporal resolution across the cycle.}
    \label{fig:irad_cycles}
\end{figure}

Figure~\ref{fig:irad_cycles} shows an example of mass-transfer cycles for a representative model with $\alpha_\mathrm{acc} = 0.1,\, \alpha_\mathrm{irr} = 0.75,\, \alpha_\mathrm{wind} = 0.001,\, \beta = 0.55$. Immediately after the onset of mass transfer, the system exhibits low-amplitude oscillations without complete detachment. At later times the cycles occur more rarely and the variability becomes more pronounced with $\dot{M}$ jumping from $\lesssim10^{-20}$ to $\sim 10^{-8} \,M_\odot\,\mathrm{yr^{-1}}$. After $\approx 600\,\mathrm{Myr}$, $\dot{M}$ rises to $\gtrsim 10^{-6}\,M_\odot\,\mathrm{yr^{-1}}$ and the evolution becomes unstable. The zoomed-in plots show the morphology of individual cycles with the system spending only a small fraction of each cycle near the peak $\dot{M}$. The characteristic period of such cycles is set by the thermal adjustment of the donor's outer envelope and is therefore of order a few Myr (see bottom panel of Fig.~\ref{fig:irad_cycles}). Such timescales are far longer than the $\sim$century observational baseline, so direct detection of a cycle in a single object is not feasible. Nevertheless, if large-amplitude cycles of this kind were common, a population observed at random phases would exhibit substantial scatter in inferred $\dot{M}$, which is not seen in CVs. This suggests that strong irradiation-driven mass-transfer cycles are unlikely to occur in real CVs, although more detailed simulations are required to robustly assess their plausibility.

\subsection{Smoothing on a characteristic timescale}

We tried suppressing the rapid irradiation response by applying temporal smoothing to the absorbed irradiation power on the characteristic thermal timescale of the donor's convective envelope. For a chemically homogeneous star, the convective-envelope timescale can be approximated by (see Eq. (66) from \citealt{2004A&A...423..281B}):

\begin{equation}
\label{eq:conv_time}
\tau_\mathrm{ce} \approx \frac{3}{7}\frac{M_\mathrm{ce}}{M_2}\tau_\mathrm{KH},
\end{equation}

\noindent where $\tau_\mathrm{KH}=GM_2^2/R_2L_2$ is the donor's Kelvin-Helmholtz thermal timescale, and $M_\mathrm{ce}$ is the mass of the convective envelope. We compared the evolution timestep $dt$ with $\tau_\mathrm{ce}$ and damped the irradiation response by updating the absorbed power only partially each step:

\begin{equation}
\label{eq:smooth}
P_\mathrm{abs}^i = P_\mathrm{abs}^{i-1} + f(P_\mathrm{abs}^\mathrm{inst}-P_\mathrm{abs}^{i-1}),
\end{equation}

\noindent where $f \equiv dt/\tau_\mathrm{ce}$ is the smoothing factor, $P_\mathrm{abs}^{i-1}$ is the absorbed irradiation power from the previous simulation step, and $P_\mathrm{abs}^\mathrm{inst}$ is the instantaneous absorbed power calculated from Eq.~\ref{eq:Pabs}. In practice, we took $f = \text{min}(1,dt/\tau_\mathrm{ce})$ so that $0\le f\le1$ for all timesteps. This approach allowed us to completely suppress mass-transfer cycles as typically during accretion $dt \ll \tau_\mathrm{ce}$ and the smoothing is strong with $f\lesssim0.01$, so $P_\mathrm{abs}$ changes by at most $\lesssim1\%$ per timestep. We stress that this is an ad hoc smoothing prescription. While the heating loop (Fig.~\ref{fig:irad_feedback}) can plausibly be damped on the convective envelope timescale, it is unlikely that the wind loop should be slowed by the same mechanism and on a similarly long timescale.

\section{Evolution with the i$\tau$SBD MB model}
\label{sec:itauSBD}

\begin{figure*}[h!]
    \centering
    \includegraphics[width = 0.49\linewidth]{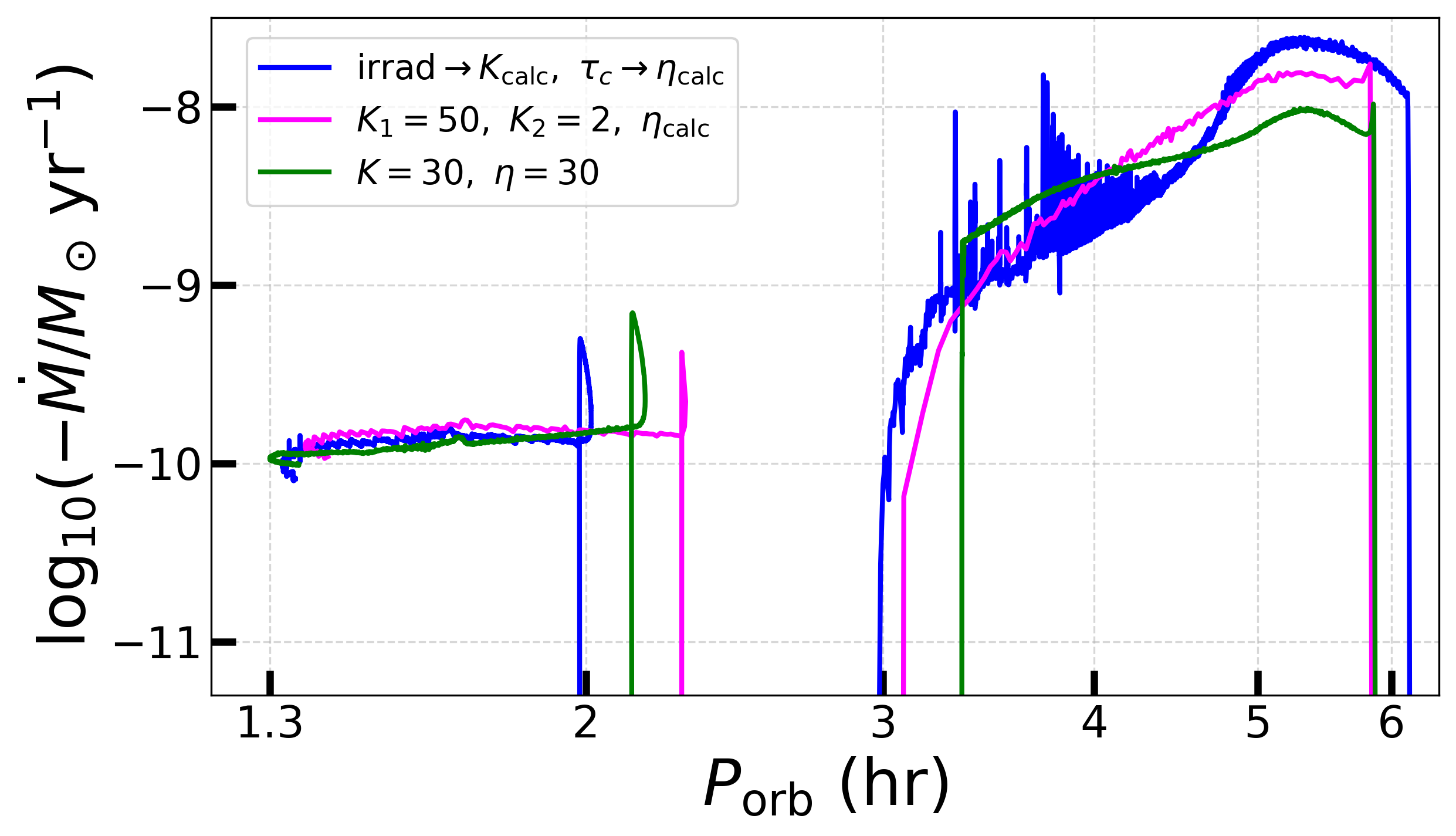}
    \includegraphics[width = 0.49\linewidth]{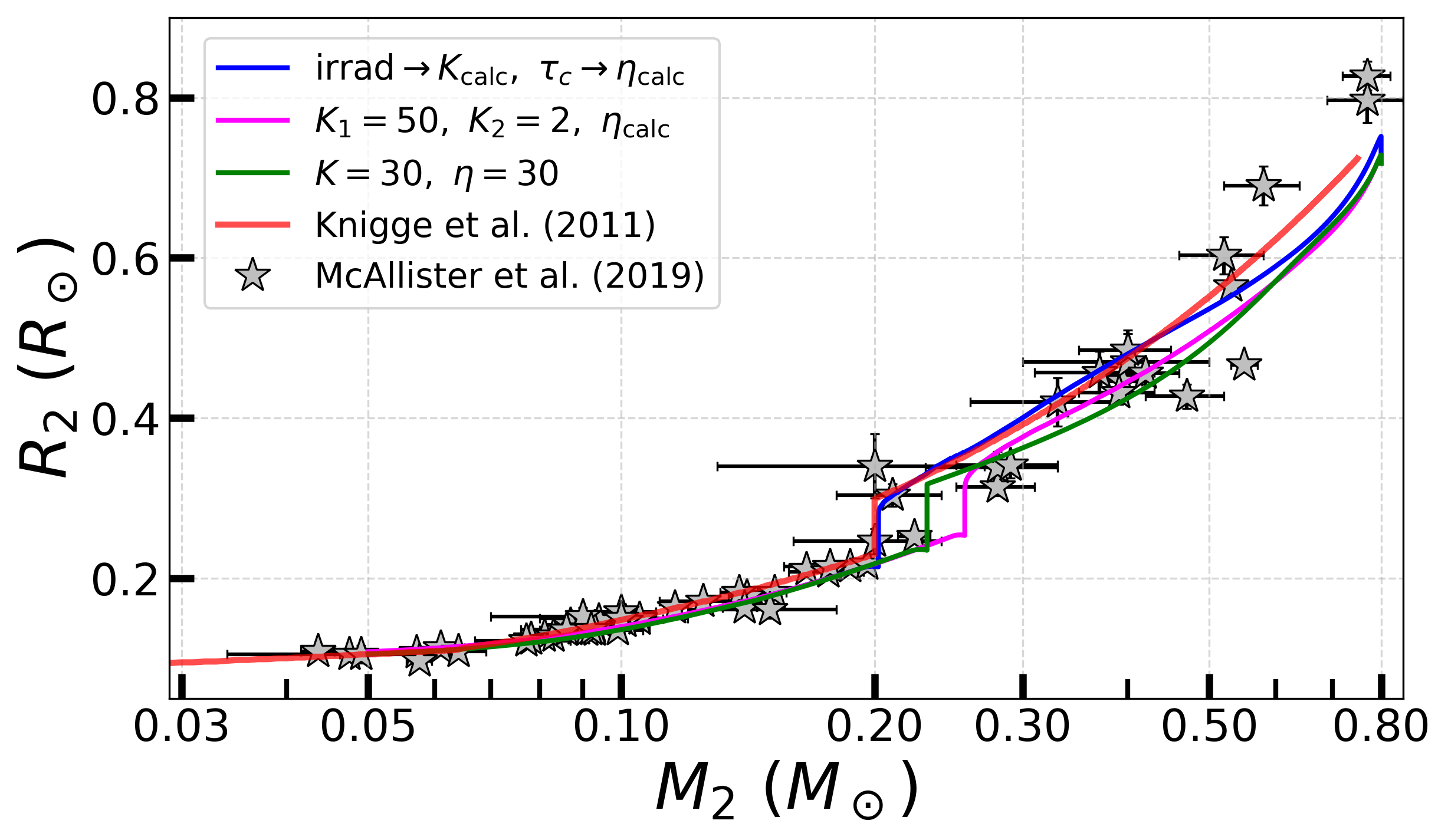}
    \includegraphics[width = 0.49\linewidth]{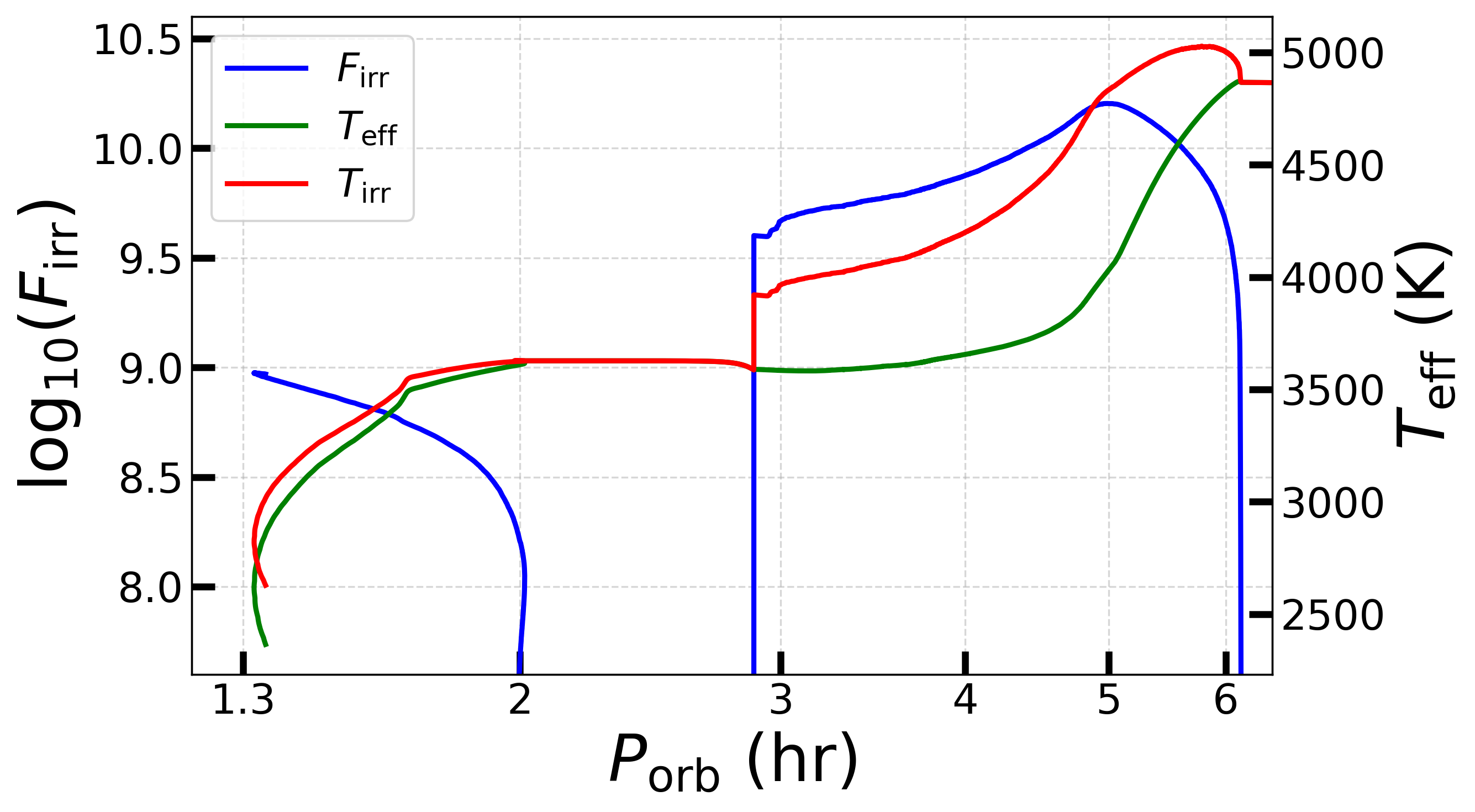}
    \includegraphics[width = 0.49\linewidth]{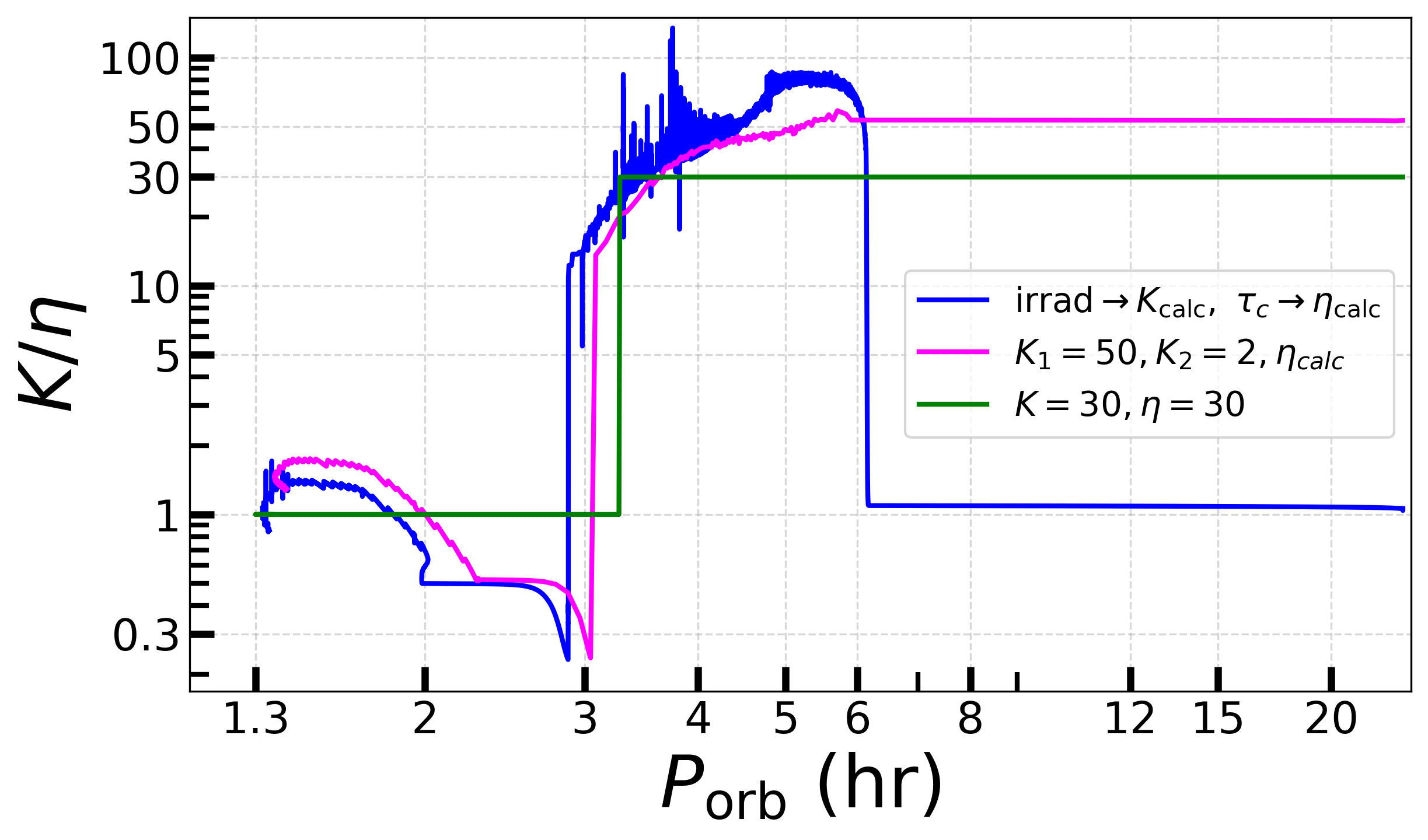}
    \caption{CV evolution with the i$\tau$SBD MB model, including irradiation-driven winds and convective turnover times computed directly from the stellar structure. The adopted irradiation parameters are $\alpha_\mathrm{acc} = 0.1,\, \alpha_\mathrm{irr} = 0.5,\, \alpha_\mathrm{wind} = 10^{-1},\, \beta = 0.40$ (with $\alpha_\mathrm{wind}$ reduced to $10^{-4}$ in the period gap). Upper panels: Same as in Fig.~\ref{fig:SBDisrupt}, with the smoothed i$\tau$SBD MB model added in blue. Lower left: Irradiation flux $F_\mathrm{irr}$ (blue), donor global effective temperature $T_\mathrm{eff}$ (green), and effective temperature of the irradiated hemisphere $T_\mathrm{irr}$ (red) as functions of orbital period. Lower right: Ratio of the boost and disruption parameters $K/\eta$ as a function of orbital period for the three models.}
    \label{fig:irad_result}
\end{figure*}

We refer to the combined SBD MB prescription in which the disruption is set by a structure-based $\tau_c$ calculation and the boost arises from irradiation-driven wind enhancement as the i$\tau$SBD MB model. In what follows, we adopted the temporal smoothing to suppress irradiation-induced mass-transfer cycles.

\subsection{Main results}

Fig.~\ref{fig:irad_result} presents evolutionary tracks obtained with i$\tau$SBD MB model. For comparison, we also show the structure-based $\tau_c$ model without irradiation ($K_1=50,K_2=2$) and empirical SBD MB ($K=\eta=30$). For the irradiation component we used $\alpha_\mathrm{acc} = 0.1,\, \alpha_\mathrm{irr} = 0.5,\, \alpha_\mathrm{wind} = 10^{-1},$ and $\beta = 0.40$. Once the system detaches, we reduced the wind efficiency to $\alpha_\mathrm{wind} = 10^{-4}$, analogous to the $K_1\rightarrow K_2$ transition applied in Fig.~\ref{fig:SBDisrupt}. The lower right panel in Fig.~\ref{fig:irad_result} shows how the ratio $K/\eta$ changes with orbital period for the three models. The i$\tau$SBD MB model naturally gives the required boost and disruption factors close to the empirical line ($K=\eta=30$; \citealt{2025A&A...696A..92B}). It also brings the donor mass-radius track closer than all other models to the semi-empirical donor sequence of \citet{2011ApJS..194...28K} because the donor is inflated from irradiation. Interestingly, the i$\tau$SBD MB model shows episodes of higher accretion rate in the 3--4\,hr range, which could be attributed to nova-like CVs (see Discussion next).

We also show on the lower left panel of Fig.~\ref{fig:irad_result} the irradiation flux, the donor's global effective temperature, and the effective temperature of the irradiated hemisphere as functions of orbital period. We compute the irradiation flux as the total absorbed irradiation power distributed over the effectively irradiated surface $F_\mathrm{irr}=P_\mathrm{abs}/(f_\mathrm{geom}R_2^2)$ (see also Eq.~\ref{eq:Pabs}). The temperature of the irradiated region is then estimated by adding the irradiation flux to the donor's intrinsic emergent flux, $T_\mathrm{irr} = (T_\mathrm{eff}^4 + F_\mathrm{irr}/\sigma_\mathrm{SB})^{1/4}$. The median temperature difference $T_\mathrm{irr}-T_\mathrm{eff}$ is $\approx 250\,\text{K}$, while the maximum difference reaches $\approx800\,\text{K}$ at $P_\mathrm{orb}\approx 5\,\text{hr}$, when irradiation is strongest. As expected, the accretion-driven irradiation flux vanishes in the period gap and the temperatures equalise. Below the gap, $F_\mathrm{irr}$ increases toward shorter orbital periods as the binary separation decreases.

\begin{figure}[h!]
    \centering
    \includegraphics[width = 1.0\linewidth]{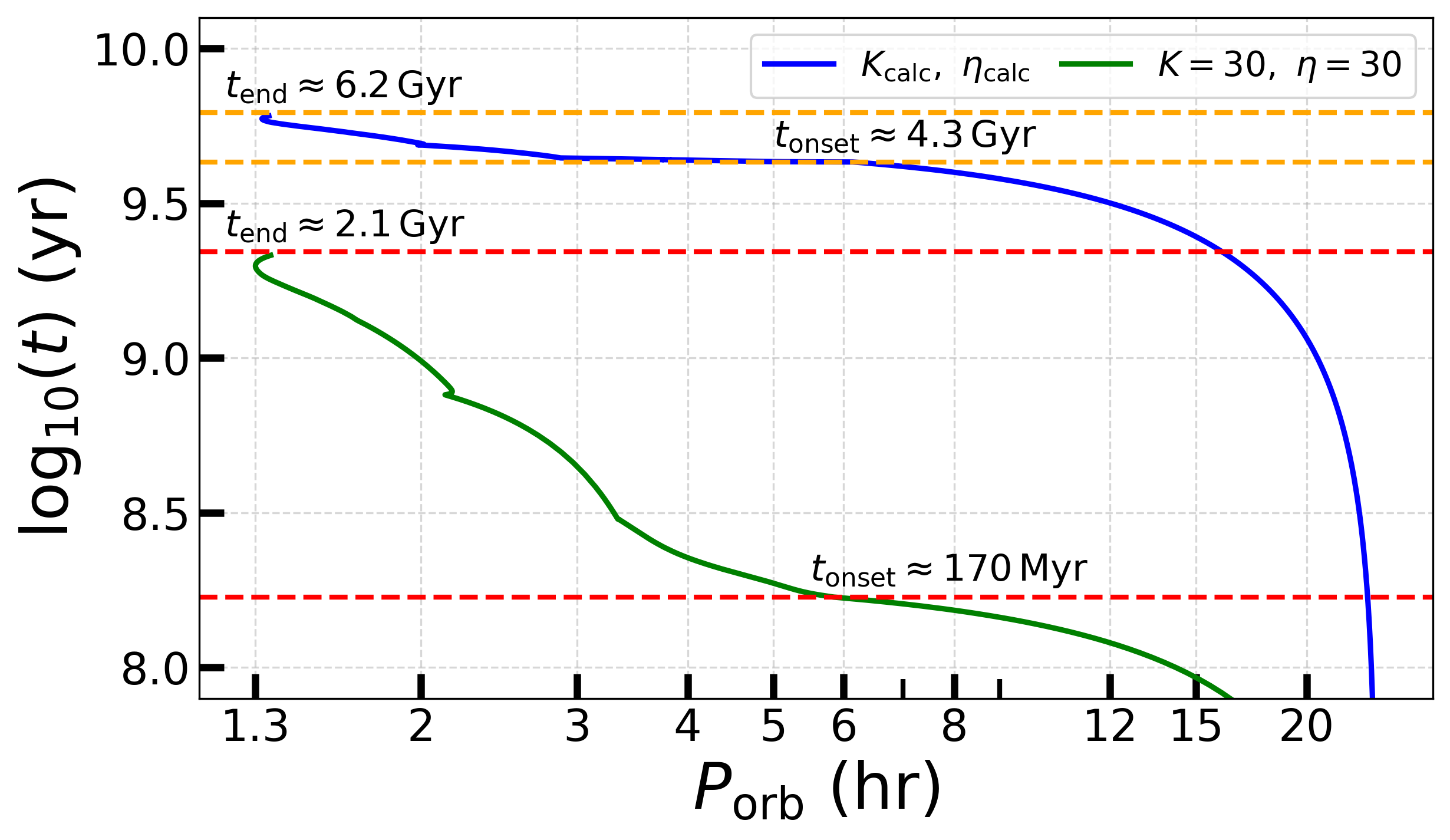}
    \caption{Evolutionary age as a function of orbital period for the i$\tau$SBD MB model (blue) and the empirical SBD MB model with $K=\eta=30$ (green). The dashed horizontal lines and annotations mark the time of first Roche lobe contact ($t_{\rm onset}$) and the termination time of the calculation when $M_2 < 0.05\,M_\odot$ ($t_{\rm end}$).}
    \label{fig:SBD_ages}
\end{figure}

An important consequence of the i$\tau$SBD MB model, compared to the empirical SBD MB calibration, is the substantially longer evolutionary timescale. In i$\tau$SBD MB, the MB boost is tied to irradiation-driven winds and therefore operates only during phases of active accretion so the post common envelope detached phase is long. By contrast, the empirical SBD MB prescription applies a constant boost ($K=30$) from the very start, which accelerates orbital shrinkage and significantly shortens the time to contact (see also lower right panel of Fig.~\ref{fig:irad_result}.) This difference is illustrated in Fig.~\ref{fig:SBD_ages}: the empirical model reaches the onset of mass transfer at $\approx170$\,Myr and terminates after $\approx2.1$\,Gyr, whereas i$\tau$SBD MB reaches contact at $\approx4.3$\,Gyr and terminates after $\approx6.2$\,Gyr. Such differences may be important when confronting models with observables that depend on long-term thermal evolution of the WD. For example, the time to the onset of mass transfer should not be shorter than the typical WD pre-contact cooling time inferred for detached CV progenitors \citep{2003A&A...406..305S, 2011A&A...536A..42Z}. It is also increasingly argued that the generation of strong magnetic fields in WDs requires timescales of order a few Gyr (see \citealt{2021NatAs...5..648S, 2025A&A...698L..22S} and references therein). While the i$\tau$SBD MB sequence yields a comparatively long pre-contact time of $t_\mathrm{onset}\approx4.3\,$Gyr, it should be kept in mind that we adopted an initial orbital period of $1\,$d. The observed period distribution of post common envelope binaries suggests shorter periods about $0.5\,$d \citep{2011A&A...536A..43N, 2011A&A...536L...3Z}, which would reduce $t_\mathrm{onset}$ in our model accordingly. Overall, the i$\tau$SBD MB timescales are therefore more consistent with expected CV formation ages.

\subsection{Model behaviour for different parameters}

\begin{figure*}[h!]
    \centering
    \includegraphics[width = 0.49\linewidth]{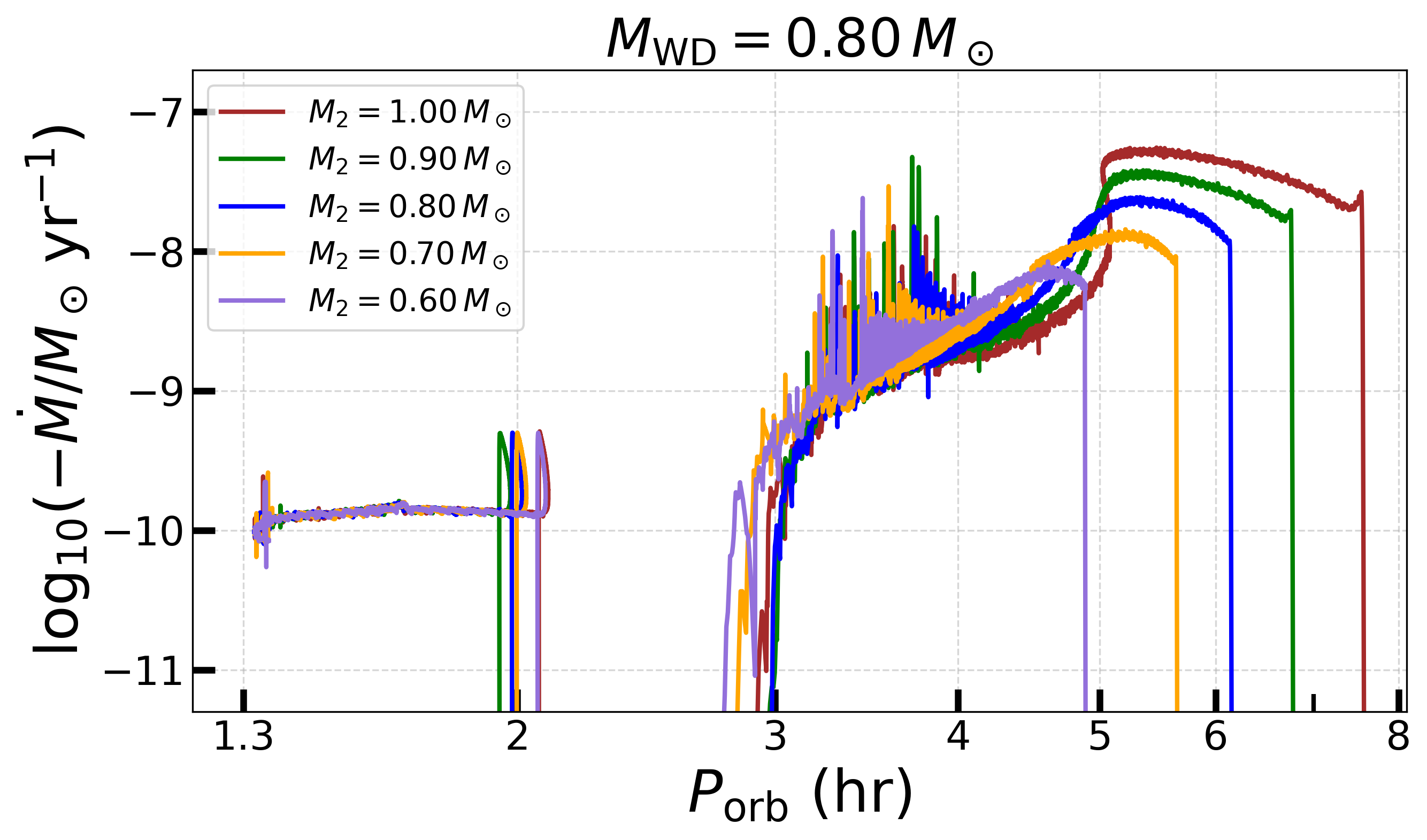}
    \includegraphics[width = 0.49\linewidth]{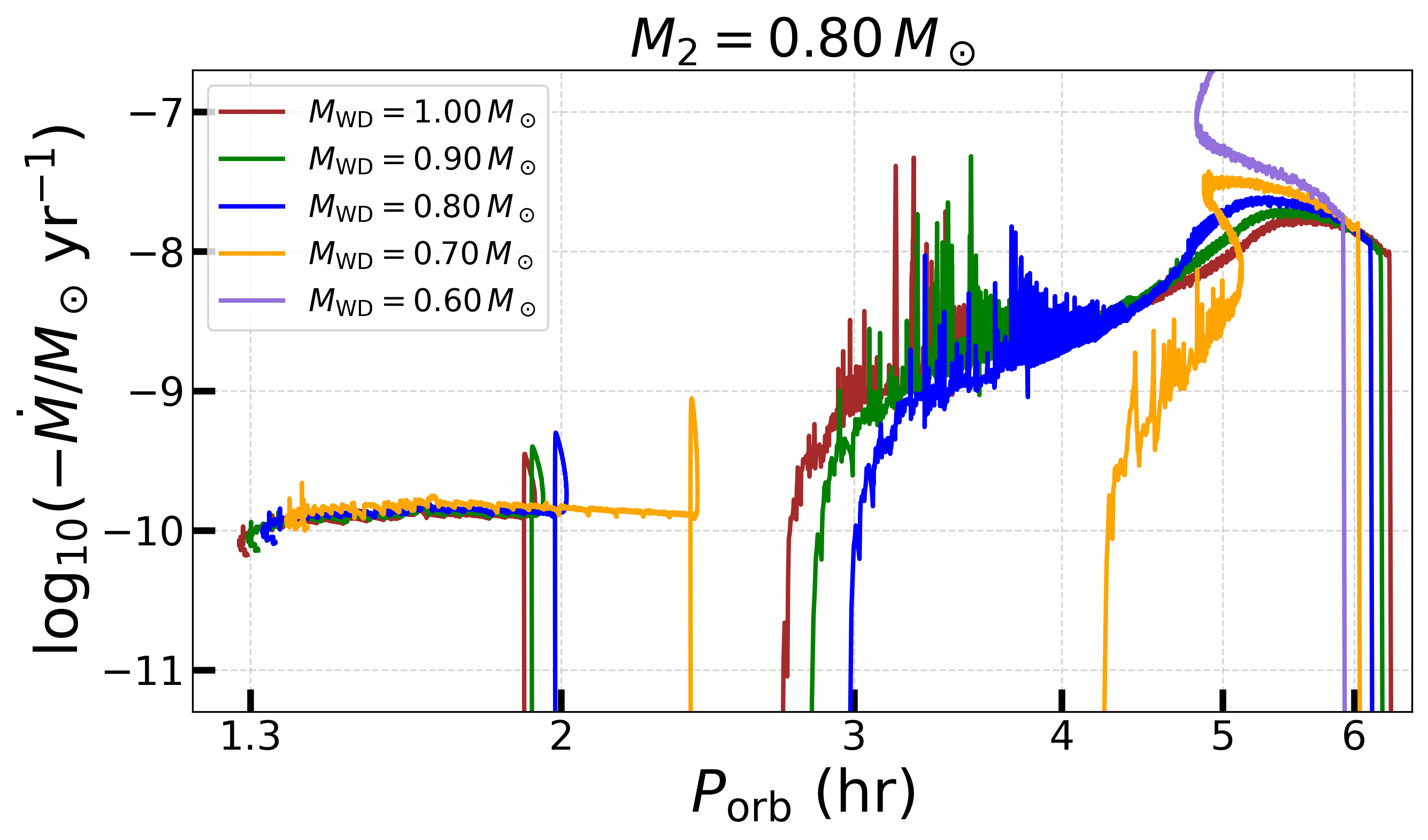}
    \includegraphics[width = 0.49\linewidth]{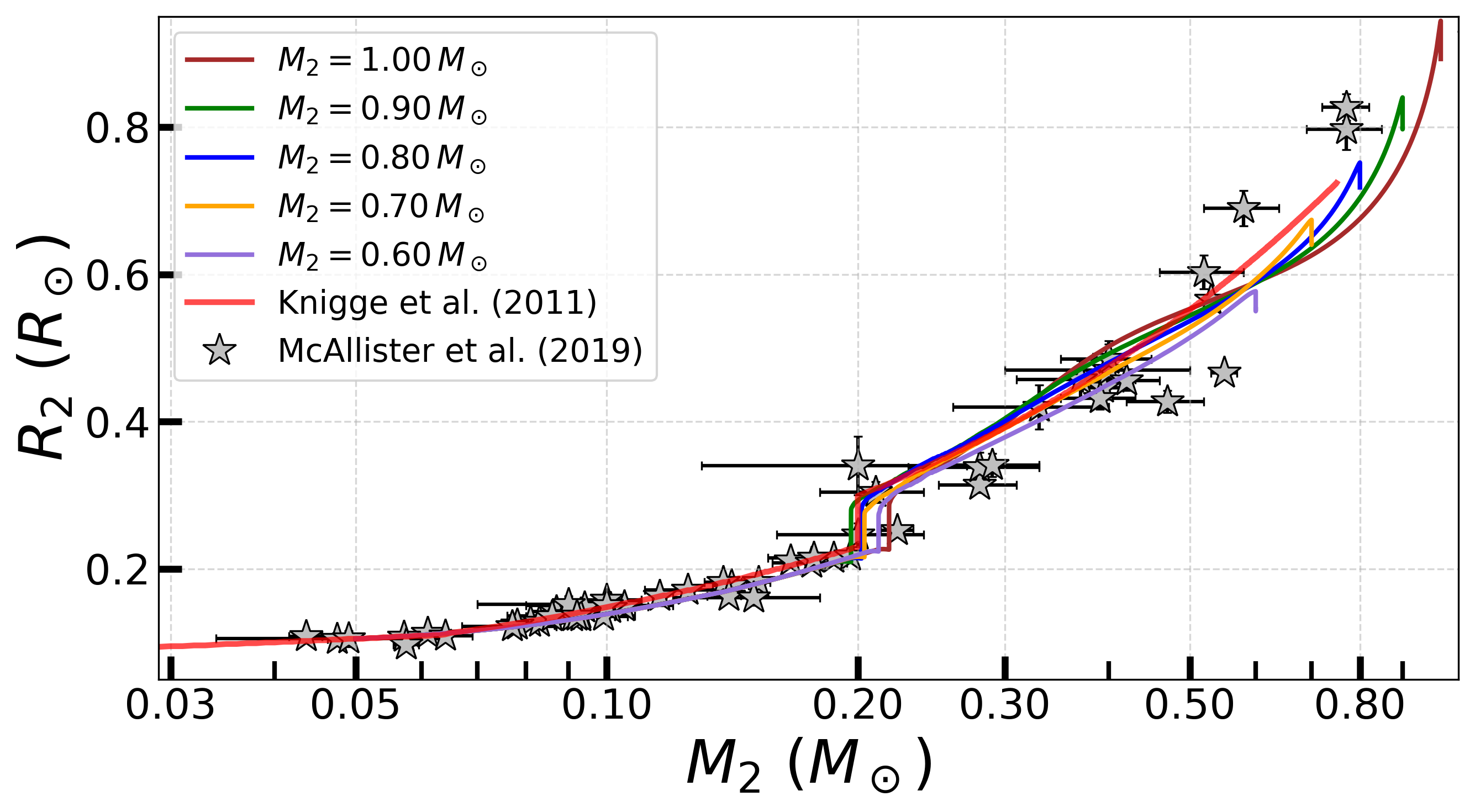}
    \includegraphics[width = 0.49\linewidth]{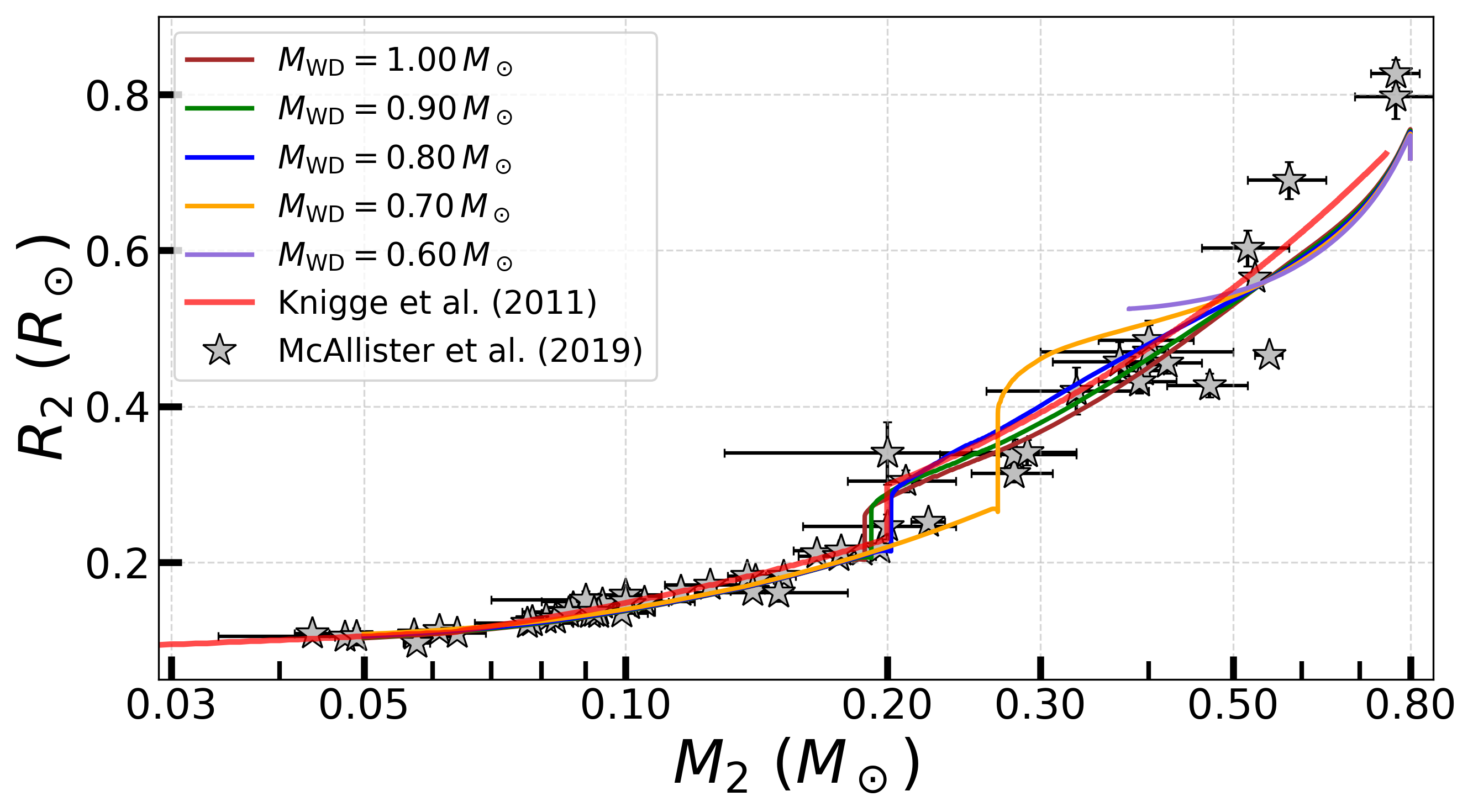}
    \caption{Evolutionary tracks for different component masses using the i$\tau$SBD MB model. $\dot{M}$ versus $P_\mathrm{orb}$ (top row) and $R_2$ versus $M_2$ (bottom row). Left column: tracks computed for fixed $M_\mathrm{WD}=0.8\,M_\odot$ while varying the initial donor mass. Right column: tracks computed for fixed $M_2=0.8\,M_\odot$ while varying the initial WD mass. The blue curve corresponds to the fiducial model from Fig.~\ref{fig:irad_result}. Other plot elements are the same as in Fig.~\ref{fig:SBDisrupt}.}
    \label{fig:irad_masses}
\end{figure*}

\begin{figure}[h!]
    \centering
    \includegraphics[width = 1.0\linewidth]{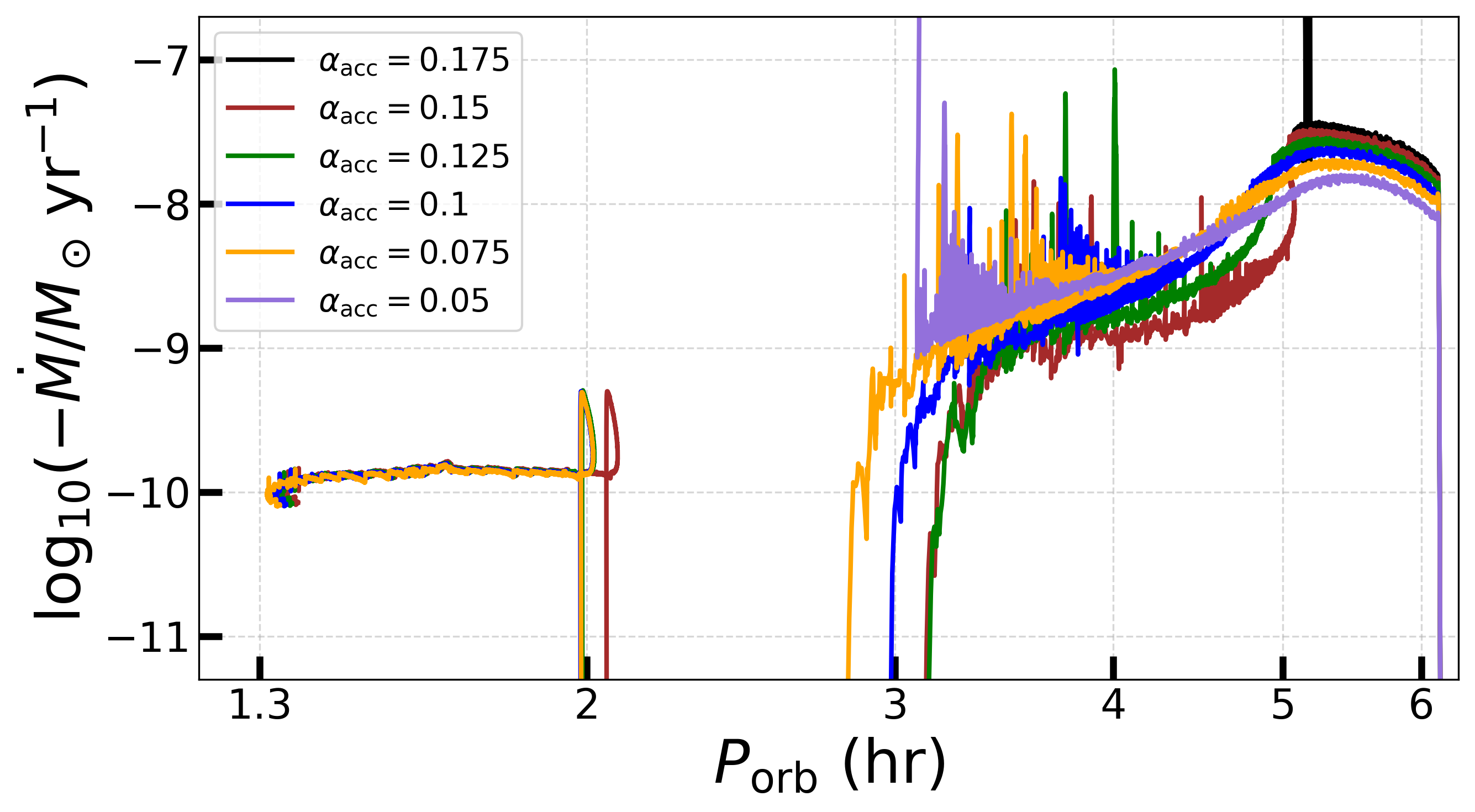}
    \includegraphics[width = 1.0\linewidth]{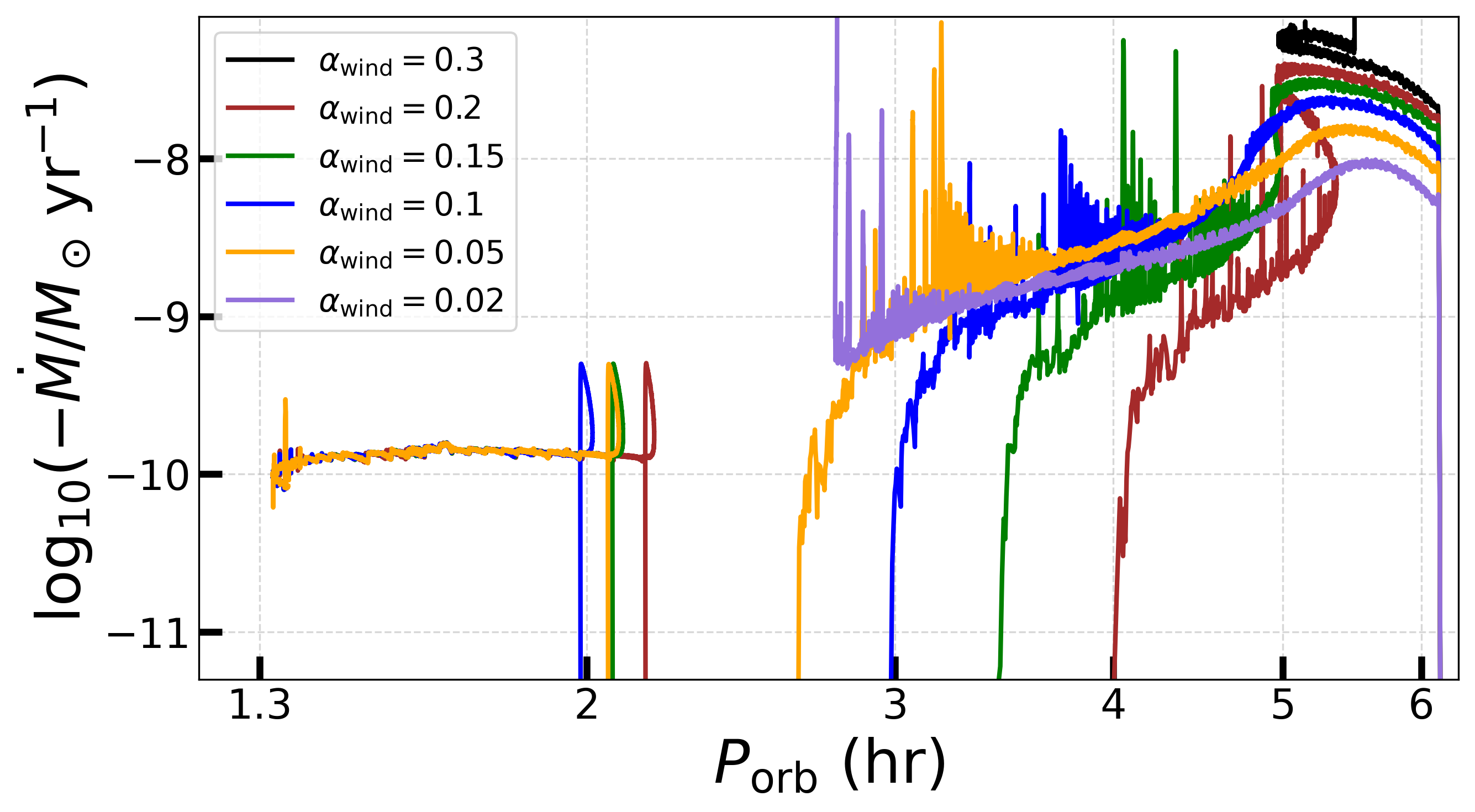}
    \caption{Dependence of the i$\tau$SBD MB evolutionary tracks in the $\dot{M}\text{--}P_\mathrm{orb}$ plane on model parameters. The top panel varies $\alpha_\mathrm{acc}$ and the bottom panel varies $\alpha_\mathrm{wind}$, with all other parameters fixed. The blue curve corresponds to the fiducial model from Fig.~\ref{fig:irad_result}.}
    \label{fig:irad_params}
\end{figure}

Fig.~\ref{fig:irad_masses} shows simulations with the i$\tau$SBD MB model using different initial WD and donor masses. Overall, the evolutionary tracks are broadly consistent, showing small shifts in the period-gap boundaries and in the location of the period minimum. As a consequence of using eCAML, systems with low WD masses ($M_\mathrm{WD}\lesssim0.6\,M_\odot$) reach dynamically unstable mass transfer and are expected to merge \citep{2016MNRAS.455L..16S}. The simulation with $M_\mathrm{WD}=0.7\,M_\odot$ also shows an atypical evolution, with the onset and end of the period gap occurring at longer periods than in the neighbouring models. Compared to the empirical SBD MB model \citep{2025A&A...696A..92B}, the mass dependence is different because irradiation introduces an additional mass-dependent feedback on the secular evolution (see Eqs.~\ref{eq:Lx},~\ref{eq:Pabs}, and \ref{eq:Mdot_wind}). It was emphasized before that the stability of irradiation prescriptions is sensitive to the assumed CAML strength \citep{2000A&A...360..969R}. This raises the possibility that irradiation may contribute to destabilising mass transfer even for weaker eCAML than assumed here. A detailed exploration of the interplay between irradiation feedback and CAML is beyond the scope of this work.

The main uncertain parameters of the i$\tau$SBD MB model are the efficiency factors $\alpha_\mathrm{acc}$, $\alpha_\mathrm{irr}$, $\alpha_\mathrm{wind}$. In our framework, $\alpha_\mathrm{acc}$ and $\alpha_\mathrm{irr}$ are mathematically degenerate and therefore affect the secular evolution in the same way. Figure~\ref{fig:irad_params} shows the resulting changes in the evolutionary tracks in the $\dot{M}\text{--}P_\mathrm{orb}$ plane when varying $\alpha_\mathrm{acc}$ and $\alpha_\mathrm{wind}$. Since $\alpha_\mathrm{acc}$ enters both the heating and wind feedback loops, whereas $\alpha_\mathrm{wind}$ affects only the wind loop, the model is more sensitive to $\alpha_\mathrm{acc}$. For both too low and too high efficiencies the system enters runaway mass transfer. For sufficiently low efficiencies, unstable mass transfer is triggered during an episode of enhanced $\dot{M}_\mathrm{acc}$ as the donor approaches the fully convective boundary. For high efficiencies, the simulations show a phase of reversed orbital-period evolution around $P_\mathrm{orb}\approx5\,$hr, in which the system widens and exhibits $\dot{P}>0$ in contrast to the usual $\dot{P}<0$. The model is even more sensitive to the exponent $\beta$ that sets how the MB torque scales with the donor wind. Increasing it from $\beta=0.4$ to $\beta=0.5$ at fixed values of the other parameters is sufficient to drive the system into unstable mass transfer.

\section{Discussion}
\label{sec:discussion}

\subsection{CV orbital period gap and minimum}

An observational requirement for CV evolutionary models is to reproduce the orbital period gap and the period minimum. In the classical disrupted MB picture, the gap was explained by assuming that MB drops sharply once the donor becomes fully convective, causing the system to detach temporarily \citep{1983ApJ...275..713R, 1983A&A...124..267S}. In practice, CV evolution above the gap is usually described with smooth MB prescriptions such as the RVJ law, while the disruption is set explicitly by sharply reducing or switching off MB once the donor becomes fully convective \citep{1993A&A...271..149K, 2001ApJ...550..897H, 2011ApJS..194...28K, 2018MNRAS.478.5626B, 2025A&A...696A..92B}. However, in this form the prescription is essentially phenomenological because the disruption is imposed by hand rather than emerging from the donor structure itself. Very few physically motivated prescriptions produce a period gap without such an ad hoc MB shutdown. One notable example is the double-dynamo model of \citet{2022MNRAS.513.4169S}. Our results suggest an alternative route in which the key trigger is the donor structure dependence of $\tau_c$.

Our i$\tau$SBD MB model produces a gap in the canonical 2--3\,hr interval and a period minimum near $\approx 1.3\,\text{hr}=78\,$min (see Figure~\ref{fig:irad_result}). To reproduce the mass-transfer rates below the period gap and the location of the period minimum our model requires an explicit change of the boost $K$ across the gap (see Figure~\ref{fig:SBDisrupt}). In our framework, the most plausible parameters for such a transition are the wind-related quantities $\alpha_\mathrm {wind}$ and/or $\beta$, which are expected to depend on the donor's magnetic field morphology. Physically, this may reflect a transition to a more complex magnetic topology as the donor crosses the fully convective boundary. Since the wind braking efficiency depends on the magnetic field morphology through the amount of open magnetic flux, more complex fields are expected to produce weaker AML \citep{2015ApJ...798..116R, 2016A&A...595A.110G, 2018ApJ...862...90G}. In this context, \citet{2026A&A...708L..11B} found that in a different saturated MB formulation only a modest suppression of MB, by a factor of $\sim$ 2--3, is sufficient to reproduce the period gap. This suggests that the stronger disruption required in the original SBD MB framework may be partly overestimated by the underlying prescription, whereas our $\tau_c$ spike still naturally produces a moderate disruption of order $\eta\sim5$.

\begin{figure}[h!]
    \centering
    \includegraphics[width = 1.0\linewidth]{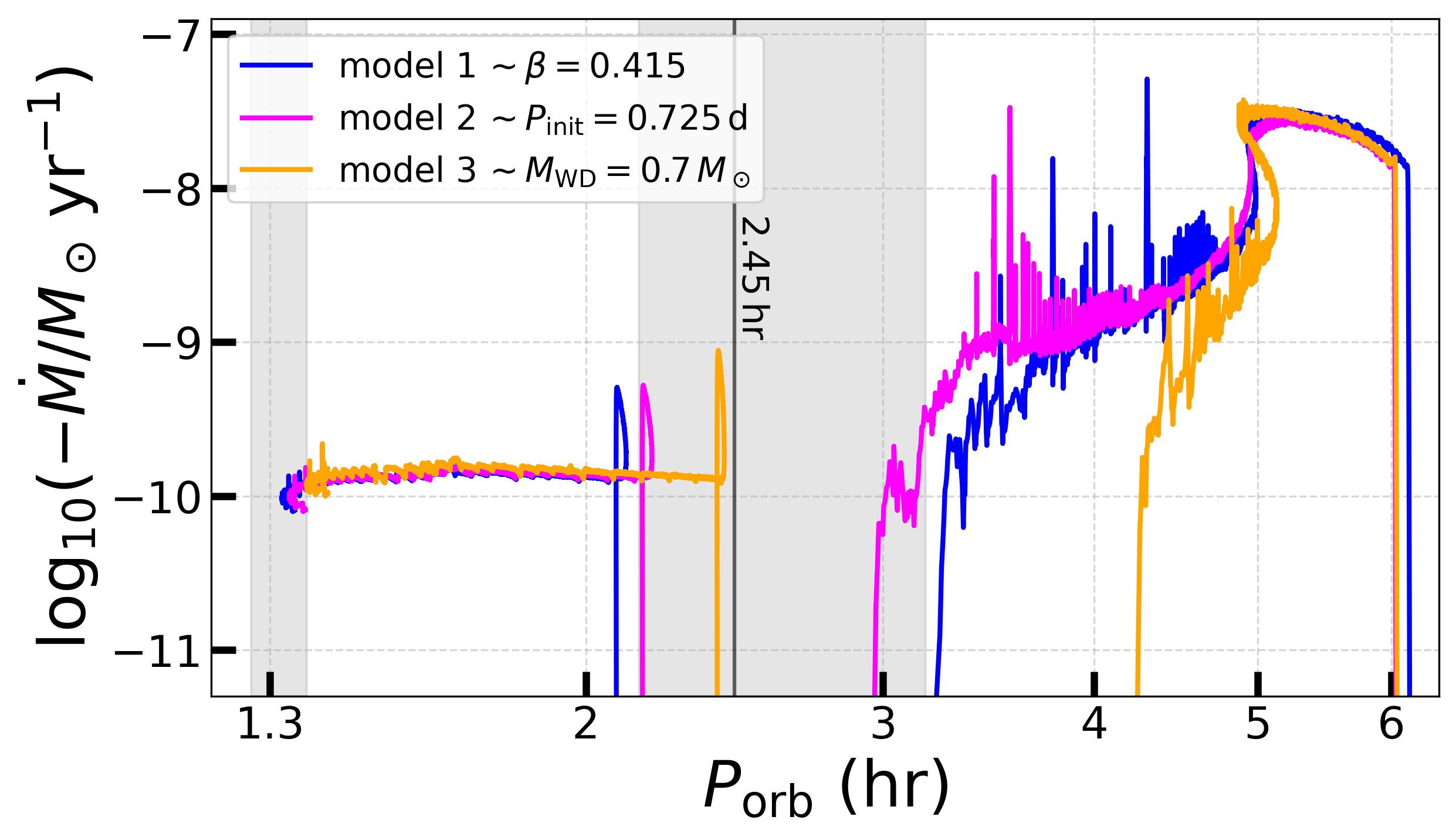}
    \caption{$\dot{M}$ versus $P_\mathrm{orb}$ for the calibrated models used to reproduce the CV period gap and period minimum (see text for the model parameters). The approximate period minimum at 76--82\,min and the period gap at 2.15--3.18\,hr are shown by the grey shaded regions \citep{2006MNRAS.373..484K}. The vertical black line marks the recently revised lower edge of the period gap at 2.45\,hr \citep{2024A&A...682L...7S}. Models 1 and 2 were calibrated to match the classical gap, while model 3 shows that a lower initial WD mass of $0.7\,M_\odot$ helps reproduce the revised lower edge.}
    \label{fig:irad_calib}
\end{figure}

The period gap and minimum have been refined as larger and more homogeneous samples became available. Some of the widely adopted values place the period gap at 2.15--3.18\,hr, and the period minimum at 76--82\,min \citep{2006MNRAS.373..484K, 2009MNRAS.397.2170G, 2019MNRAS.486.5535M}. We tried to reproduce more accurately the observed period gap and period minimum (models 1 and 2 on Figure~\ref{fig:irad_calib})\footnote{The simulation outputs for the calibrated models presented here are available on Zenodo as machine-readable tables.}. We adopted as a starting point the irradiation parameters $\alpha_\mathrm{acc} = 0.1,\, \alpha_\mathrm{irr} = 0.5,\, \alpha_\mathrm{wind} = 10^{-1}$ (reduced to $10^{-4}$ in the period gap), and $\beta = 0.40$. For model~1, we found that increasing $\beta$ to $0.415$ shifts the gap edges from 2--3\,hr to longer periods at $\approx2.1\text{--}3.2\,$hr. We then found that the initial orbital period also affects the width of the gap. For model~2, we used $P_{\rm init}=0.725\,\mathrm{d}$ together with $\alpha_{\rm acc}=0.13$, which provides a better match to the lower edge of the period gap.

More recently, \citet{2024A&A...682L...7S} suggested a revised lower edge of the period gap at $2.45\,\mathrm{hr}$. In our models, the WD mass affects the location of this lower edge. As shown by model 3 in Figure~\ref{fig:irad_calib}, an initial WD mass of $0.7\,M_\odot$ provides a better match to the revised value (see also Figure~\ref{fig:irad_masses}). The SDSS~I--IV sample used by \citet{2024A&A...682L...7S} to infer the period gap edges is the largest homogeneous CV sample currently available \citep{2023MNRAS.524.4867I}. One possible interpretation is that the refined lower edge reflects the inclusion of systems hosting relatively low mass WDs, below the mean value of $\langle M_\mathrm{WD} \rangle \approx 0.8\,M_\odot$ \citep{2011A&A...536A..42Z,2022MNRAS.510.6110P}. In this picture, our model suggests that the 2.15--2.45\,hr interval may be populated preferentially by CVs with lower mass WDs, which could now be revealed in the SDSS sample.

\subsection{Implications for the accretion rates of nova-like CVs}

\begin{figure*}[h!]
    \centering
    \includegraphics[width = 0.49\linewidth]{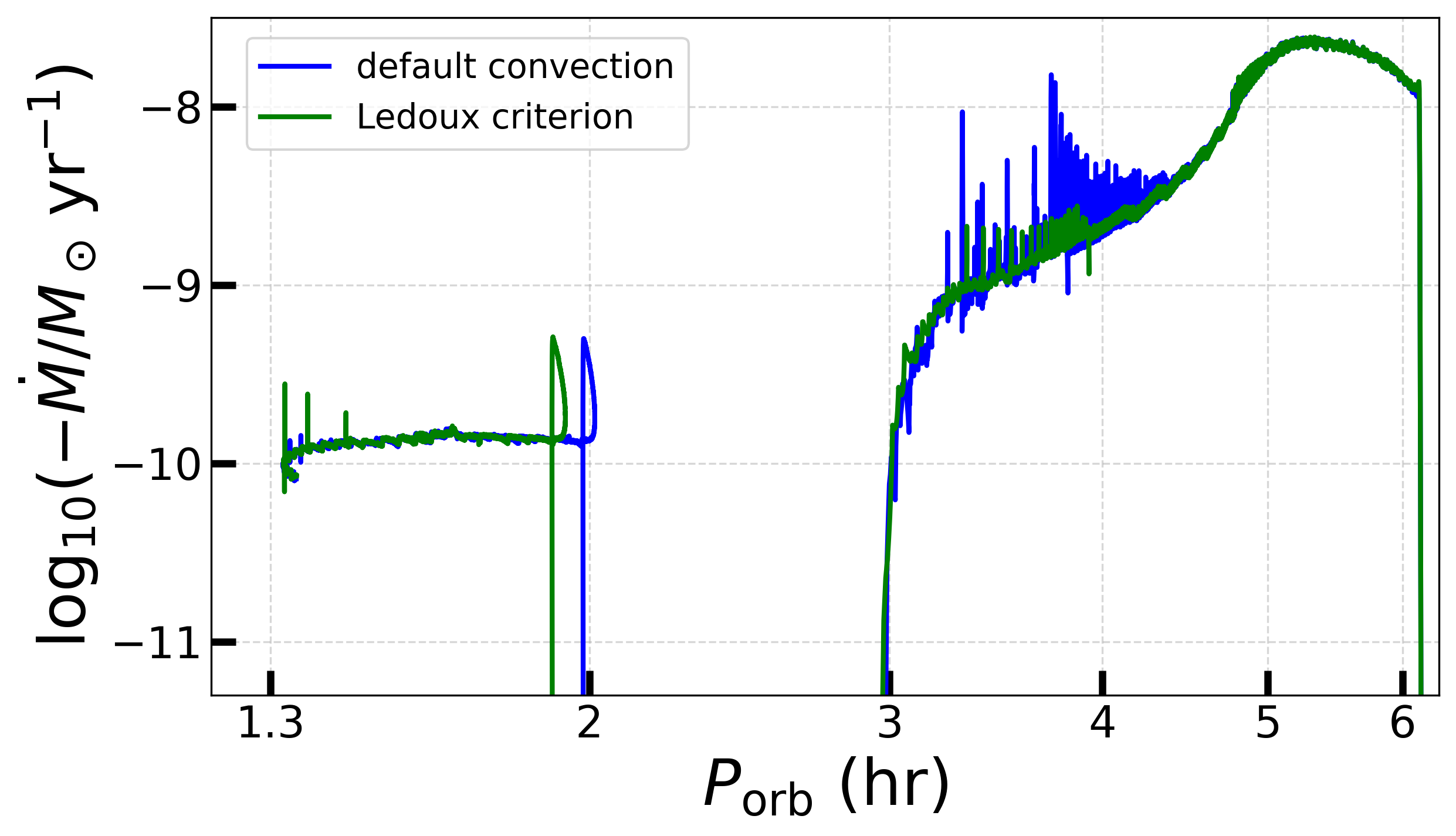}
    \includegraphics[width = 0.49\linewidth]{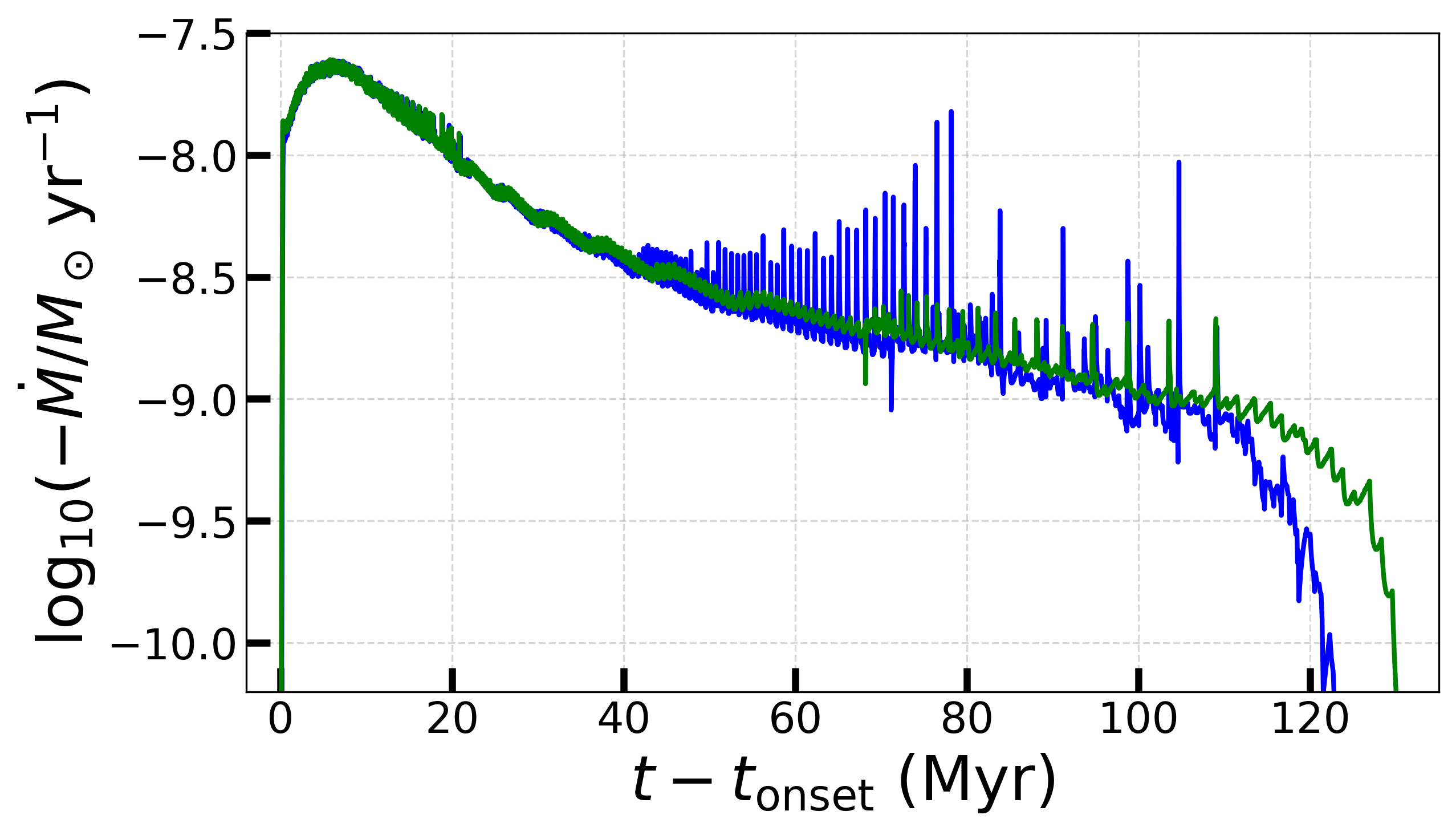}
    \includegraphics[width = 0.49\linewidth]{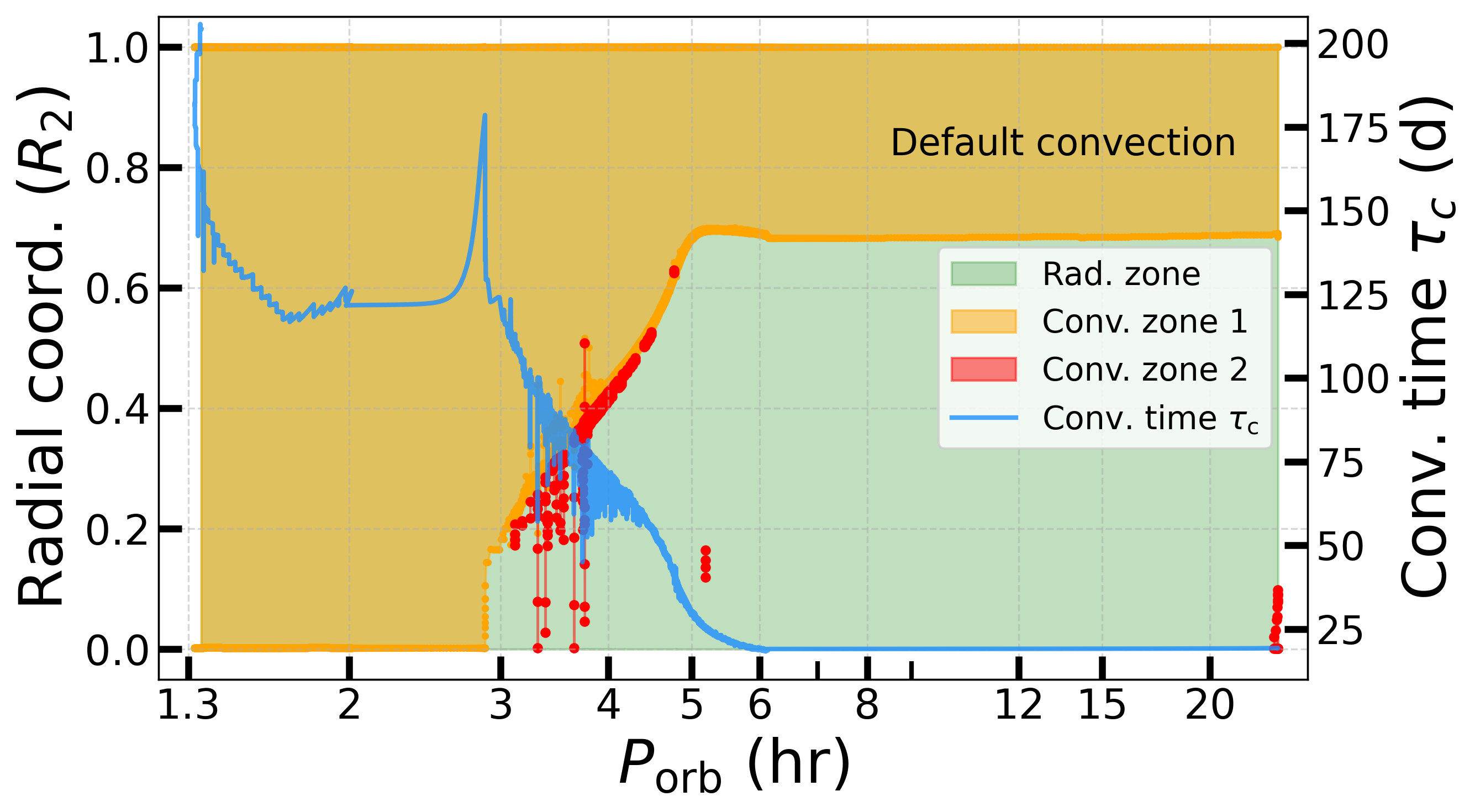}
    \includegraphics[width = 0.49\linewidth]{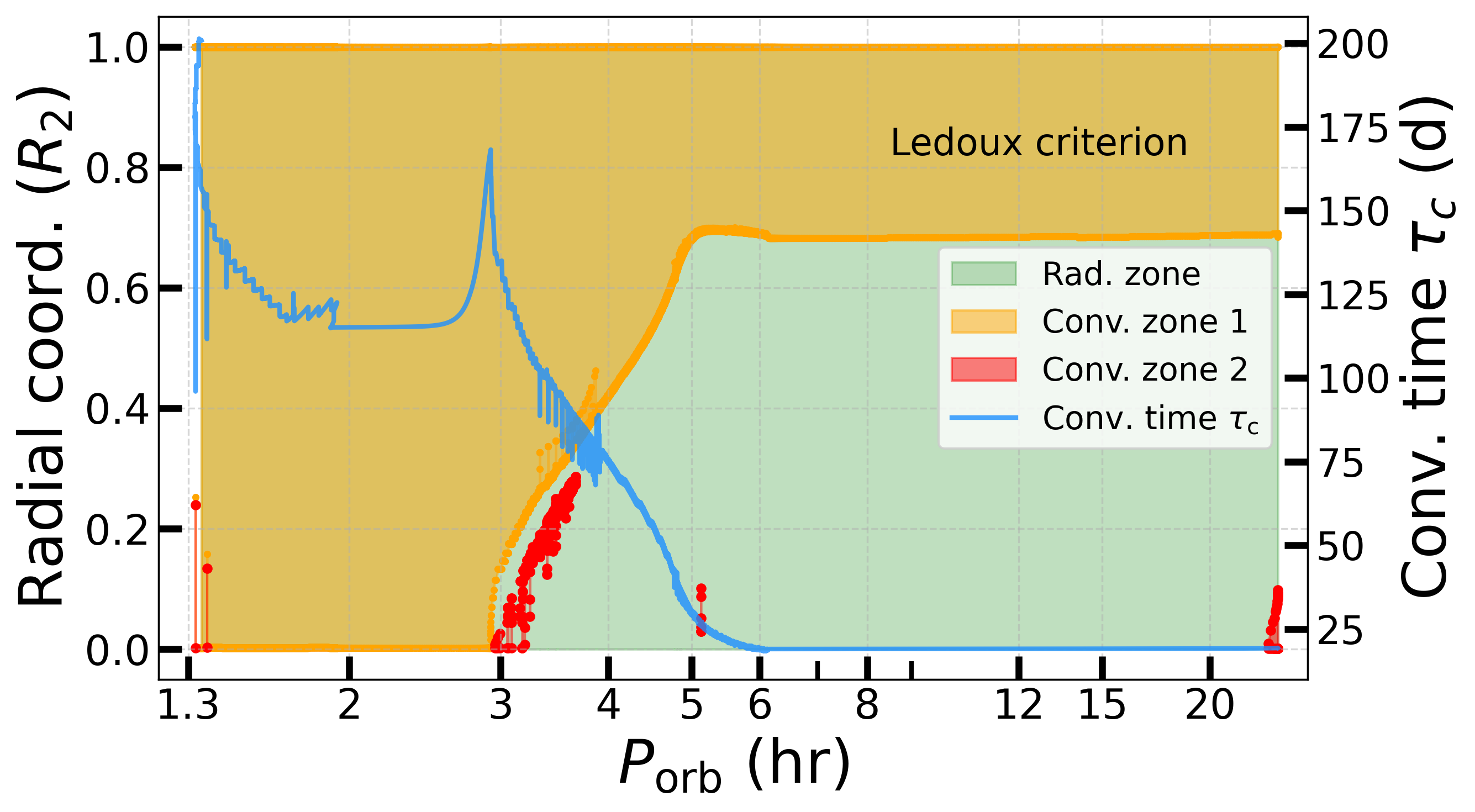}
    \caption{Simulations with the i$\tau$SBD MB model illustrating episodes of enhanced mass transfer that may be associated with nova-like CVs. Upper left: $\dot{M}$ versus $P_\mathrm{orb}$. Upper right: $\dot{M}$ versus time since the onset of mass transfer. Lower panels: evolution of convective and radiative regions, together with the corresponding convective turnover time $\tau_c$, for the default convection with Schwarzschild criterion (left) and the Ledoux criterion with additional mixing processes enabled (right).}
    \label{fig:novalikes}
\end{figure*}

Nova-like CVs are a subclass of CVs characterised by bright, persistently hot, and apparently stable accretion discs \citep{2003cvs..book.....W, 2007AJ....134.1923P}. In contrast to dwarf novae, nova-likes do not undergo disc-instability outbursts and instead sustain high accretion rates ($\dot{M}_\mathrm{acc}\sim5\times10^{-9}\,M_\odot\,\text{yr}^{-1}$) over long timescales \citep{2018A&A...617A..26D}. Observationally, most known nova-likes reside above the period gap, with a strong concentration at $P_{\rm orb}\simeq 3$--$4\,\mathrm{hr}$, where they may account for up to $\sim 50\%$ of the CV population \citep{2007MNRAS.377.1747R}. They are also among the strongest radio emitters in the CV population \citep{2015MNRAS.451.3801C, 2020MNRAS.496.2542H}. The elevated accretion rates of nova-like CVs are difficult to reconcile with standard CV evolution models employing classical AML prescriptions \citep{2024A&A...681A..83G}. Irradiation signatures have been observed in several nova-like CVs hosting very hot WDs, suggesting that irradiation may contribute to the unusually high mass-transfer rates in these systems \citep{2003ApJ...583..437A, 2007MNRAS.374.1359R}.

Notably, our i$\tau$SBD MB model produces episodes of enhanced mass transfer in the same period range where nova-likes are most commonly found (Fig.~\ref{fig:irad_result}). It is also clear that, across different initial conditions and model parameters, the full $P_\mathrm{orb}\sim3\text{--}4\,$hr interval can be populated by CVs undergoing such high states, reaching peak accretion rates as high as $\dot{M}_\mathrm{acc}\sim10^{-7}\,M_\odot\,\text{yr}^{-1}$ (see Figures~\ref{fig:irad_masses} and~\ref{fig:irad_params}). These high-$\dot{M}_{\rm acc}$ episodes coincide with changes in the donor's convective structure. As the star evolves toward full convection, additional convective regions appear closer to the interior and the convective envelope adjusts in depth. Fig.~\ref{fig:novalikes} illustrates this connection. When the convective envelope depth changes, the convective turnover time $\tau_c$ drops abruptly and the MB torque correspondingly increases. In physical terms, this behaviour may reflect a rapid reconfiguration of the large-scale magnetic field. We note that these enhanced $\dot{M}_{\rm acc}$ episodes occur only in models that include irradiation (see Fig.~\ref{fig:SBDisrupt} for comparison with a model without irradiation), indicating that irradiation plays an essential role in enabling these high states. Our simulations indicate that the typical duration of such enhanced $\dot{M}_{\rm acc}$ episodes is of order $\sim10^5\,$yr and the recurrence time is $\sim5\times$ longer than the duration. Interestingly, during these enhanced $\dot{M}_{\rm acc}$ episodes the model also exhibits positive $\dot{P}$, with characteristic values $\dot{P}\sim10^{-12}$. In classical MB prescriptions, sustained positive $\dot{P}$ is unachievable, whereas observations indicate that a substantial fraction of CVs display $\dot{P}>0$ \citep{2024ApJ...966..155S}. Such short-baseline measurements do not necessarily imply deficiencies in standard AML prescriptions \citep{2021arXiv211203779K, 2024arXiv240603948K}. Nevertheless, it is quite intriguing that our model naturally allows intervals of $\dot{P}>0$.

We also explored the sensitivity of this behaviour to the convection prescription by adopting the Ledoux criterion \citep{1947ApJ...105..305L} and enabling semiconvection, thermohaline mixing, and convective overshooting\footnote{For semiconvection, thermohaline and overshooting we used values from Table~2 of \citet{2025ApJ...988..102G}.}. In that case, the secondary convective regions form farther from the base of the convective envelope and have little impact on its depth, producing only modest $\dot{M}_{\rm acc}$ fluctuations (see right panels of Fig.~\ref{fig:novalikes}). This suggests that enhanced mixing (probably overshooting) can weaken the high-$\dot{M}_{\rm acc}$ episodes. However, the use of the Schwarzschild/Ledoux criterion requires careful evaluation depending on the star's parameters \citep{2000ApJ...534L.113C, 2014A&A...569A..63G}. A more systematic exploration will be required to determine whether, and under what conditions, such enhanced $\dot{M}_{\rm acc}$ episodes are possible.

\subsection{Implications for stellar and binary evolution}

The disruption of MB at the fully convective limit in low-mass stars is a stellar evolutionary feature, which has gained evidence from various sources. In CVs, MB disruption is directly observed as the fact of the existence of the period gap \citep{1983A&A...124..267S}. In post common envelope WD+MS binaries, the decrease in the number of systems at the fully convective limit brings evidence for MB disruption \citep{2010A&A...513L...7S, 2024A&A...682A..33B, 2026A&A...707A..76Z}. A similar effect is also found in post common envelope hot subdwarf binaries \citep{2024PASP..136l4201B}. For single low-mass stars, rotation studies show that the spin-down timescale also increases toward masses below $M_2\approx0.35\,M_\odot$ \citep{1998A&A...331..581D, 2008ApJ...684.1390R, 2011ApJ...727...56I, 2016ApJ...821...93N}. However, recent studies of single stars in the vicinity of fully convective limit indicate additional complexity. An abrupt change in the spin-down law across the fully convective boundary is possible, with braking torques instead increasing there \citep{2024NatAs...8..223L, 2024ApJ...977...15C}. Our results suggest that low-mass stars rotating in the saturated MB regime should undergo a pronounced weakening of MB upon becoming fully convective, driven by the spike in $\tau_c$.

Virtually all low-mass main-sequence stars are expected to experience a pronounced increase in the convective turnover time $\tau_c$ at the fully convective boundary \citep{2024ApJ...977...15C, 2025ApJ...988..102G}. In the SBD MB framework, which builds on the MB prescription of \citet{1995ApJ...441..865C}, this naturally results in MB disruption as $\dot{J}_\mathrm{MB}\propto\tau_c^{-2}$ in the saturated regime. However, this sensitivity to $\tau_c$ is not universal. For example, the widely used prescription of \citet{2015ApJ...799L..23M}, adopted by \citet{2026A&A...708L..11B} in their revised SBD MB model, does not retain an explicit $\tau_c$-dependence in the saturated regime. Similarly, the Convection And Rotation Boosted prescription (CARB; \citealt{2019MNRAS.483.5595V, 2019ApJ...886L..31V}), which has been applied successfully to LMXBs, yields $\dot{J}_\mathrm{MB}\propto\tau_c^{8/3}$. Positive scalings, $\dot{J}_\mathrm{MB}\propto\tau_c^{n}$ with $n>0$, typically arise when the surface magnetic field is assumed to scale with Rossby number $B\propto Ro^{-1}\propto\tau_c/P_\mathrm{rot}$. Negative scalings such as $\dot{J}_\mathrm{MB}\propto\tau_c^{-2}$ are usually tied to saturated MB formulations in which the saturation threshold satisfies $P_\mathrm{sat}\propto\tau_c$. Therefore, a spike in $\tau_c$ leads to MB disruption only in certain types of semi-empirical saturated MB prescriptions. A derivation of the MB torque that robustly predicts the saturated MB regime would therefore be highly valuable.

As shown in Section~\ref{sec:itauSBD}, our irradiation prescription can account for the enhanced MB required in CVs. The same physical mechanism should operate in other compact accreting binaries whose donors possess convective envelopes. Indeed, abnormally strong MB, interpreted as being driven by irradiation-induced winds, has been invoked to explain the evolution of several compact X-ray binaries \citep{2006MNRAS.366.1415J, 2016ApJ...830..131C, 2019ApJ...887..201X, 2026ApJ...997..162Z}. Similar effects have also been proposed for spider pulsars, where strong irradiation of the companion can drive an ablated wind and enhance MB \citep{2020MNRAS.495.3656G, 2021MNRAS.500.1592G, 2023MNRAS.525.2708C}. Remarkably, \citet{2021MNRAS.500.1592G} also found that a moderate reduction of MB at the fully convective transition best explains the observed spider sample, broadly consistent with the disruption of $\eta\sim5$ found here. This agreement between independent observational constraints supports the idea of a more universal physically motivated MB prescription.

A related question is whether irradiation-driven winds can also account for the MB boost inferred in detached binaries with hot primaries \citep{2024A&A...682A..33B, 2024PASP..136l4201B}. In detached systems, irradiation is easier to assess because it is not coupled to mass-transfer feedback. Although the incident spectrum differs from the accretion-powered irradiation considered for CVs, the same energy-limited framework can still be used for an order-of-magnitude estimate of the resulting donor wind, with the spectral differences most likely reflected in the uncertain efficiency parameters of the irradiation model. We therefore made simple estimates of the boost factor $K$ expected in detached WD+MS and sdB+MS binaries by replacing the accretion-powered luminosity with the thermal luminosity of the primary, approximated as a blackbody by $L_1 = 4\pi R_1^2 \sigma_{\rm SB} T_1^4$. The corresponding MB boost in detached binaries can be written as (see Eqs.~\ref{eq:Pabs},~\ref{eq:Mdot_wind} and~\ref{eq:Kboost}):

\begin{equation}
\label{eq:K_detach}
K \approx 1+\left( \frac{\pi \alpha_{\rm wind}\alpha_{\rm irr}\sigma_{\rm SB} R_1^2 T_1^4 R_2^3}{G M_2 a^2 \dot M_{\rm wind,base}} \right)^{\beta}.
\end{equation}

\noindent For these estimates, we fixed the donor parameters at $M_2=0.5\,M_\odot$ and $R_2=0.5\,R_\odot$, since the empirical boost in detached systems appears to be most important for binaries with $M_2\gtrsim0.4\,M_\odot$ \citep{2024A&A...682A..33B, 2024PASP..136l4201B}. We used $M_1=0.6\,M_\odot,\,R_1=0.012\,R_\odot$ and $M_1=0.47\,M_\odot,\,R_1=0.20\,R_\odot$ for systems with WD and sdB primaries, respectively. We also fixed $\alpha_\mathrm{irr}=0.7,\,\beta=0.5,\,\dot{M}_\mathrm{wind,base}=0.1\,\dot{M}_\odot,$ and $P_\mathrm{orb}=0.5\,$d. We varied the poorly constrained wind efficiency between $\alpha_{\rm wind}=10^{-3}$ and $10^{-1}$ in order to illustrate the possible spread in the resulting boost factor $K$ arising from the current uncertainties in the parameters of the irradiation model. Table~\ref{tab:detached_K_estimates} summarizes the resulting representative values of $K$ for different temperatures of the primaries. These estimates suggest that the large boost $K\gtrsim50$ inferred for PCEBs by \citet{2024A&A...682A..33B} can be reached only if the WD is sufficiently hot ($T_1\gtrsim30\,000K$). By contrast, in sdB+MS binaries irradiation-driven winds can readily provide the required large boost $K\gtrsim100$, as first proposed by \citet{2024PASP..136l4201B}. This suggests that irradiation alone may be insufficient to explain the full MB boost in some detached binaries as WD+MS, and that an additional boosting mechanism may also be required in such systems. Irradiation-driven winds may also help to explain the accretion rates in wind-fed magnetic systems such as low accretion rate polars, where the observed accretion rates of order $10^{-13}\,M_\odot\,\text{yr}^{-1}$ are difficult to reconcile with the weak intrinsic winds expected from M dwarfs \citep{2005ASPC..330..137W, 2012ApJ...747..132M} as well as some super soft X-ray sources like CAL 87, where the small mass ratio and expanding orbit deviate from the standard model for such sources \citep{1998A&A...338..957V, 2015ApJ...815...17A}.

\begin{table}
\caption{Estimated irradiation-driven boost factors in detached binaries}
\label{tab:detached_K_estimates}
\centering
\renewcommand{\arraystretch}{1.2}
\begin{tabular}{lccc}
\hline\hline
System type & $T_1$ (K) & $K_{\rm low}$ & $K_{\rm high}$ \\
\hline
WD+MS  & 10\,000 & 1.4 & 5.2  \\
WD+MS  & 30\,000 & 4.7 & 38   \\
WD+MS  & 50\,000 & 11  & 105  \\
sdB+MS & 20\,000 & 30  & 290  \\
sdB+MS & 30\,000 & 66  & 650  \\
sdB+MS & 40\,000 & 116 & 1150 \\
\hline
\end{tabular}
\tablefoot{The system parameters were fixed at representative values (see text for details). The low and high estimates correspond to wind efficiencies of $\alpha_\mathrm{wind}=10^{-3}$ and $10^{-1}$, respectively.}
\renewcommand{\arraystretch}{1.0}
\end{table}

\subsection{Model limitations and future work}

While our i$\tau$SBD MB model provides a plausible route toward a physical interpretation of the empirical SBD boost and disruption, the present implementation still relies on several simplifying assumptions that should be addressed in future studies. One useful direction will be to investigate whether the physical ingredients proposed here can be incorporated into the revised SBD MB framework of \citet{2026A&A...708L..11B}.

The treatment of irradiation is highly idealised. The irradiation component introduces a number of uncertain parameters that are currently treated as constants, independent of system properties. In reality, the accretion efficiency $\alpha_\mathrm{acc}$ should depend on the WD mass and radius, as well as on boundary-layer physics. The irradiation efficiency $\alpha_\mathrm{irr}$ should depend on the incident spectrum, the donor's atmospheric properties that determine the effective albedo, and possible shadowing by the accretion disc. The wind efficiency $\alpha_\mathrm{wind}$ should depend on the hydrodynamics of wind launching. Also, together with the exponent $\beta$, they are expected to depend on the magnetic-field topology and wind-acceleration profile \citep{2015ApJ...798..116R}. A further simplification is our assumption of a constant base wind for the donor, $\dot{M}_\mathrm{wind,base}=0.1\,\dot{M}_\odot$. A physically motivated prescription for wind mass-loss applicable to both partially and fully convective stars is clearly needed. In addition, we account only for irradiation associated with accretion and neglect direct irradiation by the hot WD itself. Such UV irradiation may also heat the donor significantly and could be especially important in detached systems with hot primaries.

Several additional ad hoc simplifications were adopted in the present CV evolution calculations. The need for temporal smoothing of $P_{\rm abs}$ to suppress irradiation-driven mass-transfer cycles indicates that the present treatment is incomplete. A more realistic approach should distinguish between the thermal response time of the donor's outer envelope and the response time of the wind and MB torque to changes in irradiation. Another key limitation of the current model is that it still requires an explicit reduction of the MB boost once the system detaches and evolves below the gap. In practice, we mimic this by changing $\alpha_\mathrm{wind}$ or $\beta$ from a high to a low value within the gap. Although such a change is physically plausible, an important goal for future work is to replace these switches with a continuous and physically motivated prescription. Finally, several simplifying assumptions in the binary evolution setup were adopted. We kept the WD mass fixed, treated the WD as a point mass without thermal evolution, and adopted an empirical CAML prescription. In addition, including the suppression of MB by coupling between the donor and WD magnetospheres \citep{2020MNRAS.491.5717B} will allow to extend the framework to magnetic CVs.

\section{Conclusions}
\label{sec:conclusions}

We have developed an extension of the empirical saturated, boosted, and disrupted magnetic braking model (SBD MB) that incorporates irradiation-driven winds and convective turnover times computed directly from the stellar structure (the i$\tau$SBD MB model). We tested this prescription for CV evolution using MESA and found that:

\begin{enumerate}
    \item Computing the convective turnover time $\tau_c$ directly from the donor structure produces a pronounced spike as the donor approaches full convection. In a saturated MB prescription ($\dot{J}_\mathrm{MB}\propto\tau_c^{-2}$), this naturally yields a rapid reduction of the braking torque and can offer a physical explanation for MB disruption in CVs.
    \item Accretion-powered irradiation can substantially modify the donor's outer layers and strongly heat the donor-facing hemisphere. Irradiation-driven winds can enhance the effective MB torque during phases of active accretion and may provide a physical origin for the MB boost in CVs. The main limitation is that the present irradiation treatment remains highly simplified, relying on uncertain efficiencies, an assumed base wind for the M-dwarf donor, and ad hoc prescriptions for smoothing the irradiation response and for reducing the boost once the donor becomes fully convective.
    \item Combining structure-based $\tau_c$ with irradiation-driven winds in the SBD MB framework yields evolutionary tracks that reproduce the main qualitative features of the CV population, including the period gap and period minimum. The model also brings the donor mass-radius relation closer to the semi-empirical donor sequence by allowing moderate donor inflation, predicts longer pre-contact evolutionary timescales that are more compatible with expected CV formation ages, and might produce episodic high-$\dot{M}_\mathrm{acc}$ episodes in the 3--4\,hr range, where nova-like CVs are concentrated.
\end{enumerate}

Overall, the i$\tau$SBD MB model provides a plausible physical interpretation of both the empirical boost and disruption factors and may be applicable to other close binaries with convective donors. The $\tau_c$ spike at the fully convective boundary offers a generic route to MB disruption in saturated-torque formulations, while irradiation-driven winds are expected to be relevant in compact binaries with sustained accretion and/or hot primaries. A key next step is to replace the present simplified irradiation prescription with a more physical treatment of wind launching and its coupling to the magnetic torque.

\section*{Data availability}

The MESA code needed to reproduce the simulations presented in this work is available on Zenodo at \url{https://zenodo.org/records/20201258}.

\begin{acknowledgements}
VD thanks Nanjing University for its hospitality during the visit and acknowledges support from Kazan Federal University. This work was supported by the National Key Research and Development Program of China (2021YFA0718500) and the Natural Science Foundation of China under grant numbers 12121003 and 12373034. IG and AS acknowledge support from Kazan Federal University. The authors thank the anonymous referee for their thoughtful review that helped significantly improve the manuscript.
\end{acknowledgements}

%
%
\bibliographystyle{aa}
\bibliography{SBD_MB}

\end{document}